\documentclass[%
 reprint,
 prm,amsmath,amssymb,
 aps,
]{revtex4-2}

\usepackage{float}
\usepackage{graphicx}
\usepackage{dcolumn}
\usepackage{bm}
\usepackage{url}
\usepackage{amsmath}
\usepackage{amssymb}
\usepackage{multirow}
\usepackage{array}
\usepackage{ulem}

\usepackage[usenames,dvipsnames]{color}

\begin{document}


\title{High-throughput screening of charge-order-induced ferroelectrics}

\author{Jose Cuevas-Medina$^{(1)}$,  Yubo Qi$^{(2)}$, Nata\v sa Stoji\' c$^{(3)}$ and Sebastian E. Reyes-Lillo$^{(4)}$}
\email{sebastian.reyes@unab.cl}

\affiliation{
(1) Doctorado en Ciencias F\'isicas, Universidad Andres Bello, Santiago 837-0136, Chile  \\
(2) Department of Physics, University of Alabama at Birmingham, Birmingham, Alabama 35233, USA \\
(3) Abdus Salam International Center for Theoretical Physics, Trieste, Italy \\
(4) Departamento de F\'isica y Astronom\'ia, Universidad Andres Bello, Santiago 837-0136, Chile
}\date{\today}

\begin{abstract}
Charge-order-induced ferroelectrics display important technological applications in spintronics devices due to the possibility of magnetoelectric coupling and fast electronic switching. However, the list of known charge-order-induced ferroelectrics remains limited, hindering the fundamental understanding of the phenomena and the optimization of materials for real applications. In this work, we develop a high-throughput workflow to screen for charge-order-induced ferroelectrics in The Materials Project database.  We use the local symmetry and bond valence sum to determine 147 materials displaying a coexistence of charge order and ferroelectric polarization. Then, ab initio simulations are used to identify 21 charge-order-induced ferroelectric candidates in which the ferroelectric polarization originates from or is structurally coupled to the charge order. For the final 21 candidates, we use symmetry-adapted modes and first-principles calculations to determine the structural coupling term between charge order and polar distortions and compute electronic properties, respectively. 
\end{abstract}

\pacs{Valid PACS appear here}
\maketitle

\section{Introduction}

Ferroelectrics, insulating materials with a spontaneous and switchable  macroscopic polarization under an external electric field, display important applications in electronic and memory devices~\cite{Rabe-book}. Ferroelectrics displaying coupling to spin, charge, orbital, or lattice degrees of freedom are promising for technological applications due to the possibility of electric field control~\cite{VanDenBrink2008, Picozzi2009}. Among these, magnetoelectric multiferroics coupling electric and magnetic degrees of freedom have received considerable interest for spintronic devices~\cite{Khomskii2009, Spaldin2005, Fiebig2016, Spaldin2019}.

An interesting group of multiferroics includes materials in which ferroelectricity is caused by charge order (CO)~\cite{VanDenBrink2008}. In a seminal work, J. Van den Brink and D. I. Khomskii introduced the concept of CO-induced ferroelectricity as a type of ferroelectric where the polar distortion arises from the combination of CO and a centrosymmetric structural distortion~\cite{VanDenBrink2008}. In conventional perovskite ferroelectrics (e.g. BaTiO$_3$ and PbTiO$_3$), ferroelectricity arises due to hybridization and chemical bonding, and the polar distortion is associated with the freezing of an unstable $\mathbf{q} = 0$ polar mode. In contrast, in CO-induced ferroelectrics, the macroscopic polarization arises from the electronic rearrangement of the ions' charge, and the freezing of a $\mathbf{q} \neq 0$ zone-boundary mode. Since the switching mechanism of the macroscopic polarization can in principle occur through nonadiabatic electronic charge transfer rather than slow ionic motion, CO-induced ferroelectrics could enable orders-of-magnitude faster polarization switching with minimal energy dissipation, offering a transformative advantage for high-speed non-volatile memory and neuromorphic devices. Additionally, previous work has shown the interplay between CO and superconductivity~\cite{Eduardo2014, Zheng2022}, colossal magnetoresistance~\cite{Sen2007}, light-induced phenomena~\cite{Kawakami2010}, metal-insulator transitions~\cite{Alonso2000, Ovsyannikov2016} and relativistic phenomena~\cite{He2018}.

\begin{figure}[htp]
\includegraphics[width=\columnwidth]{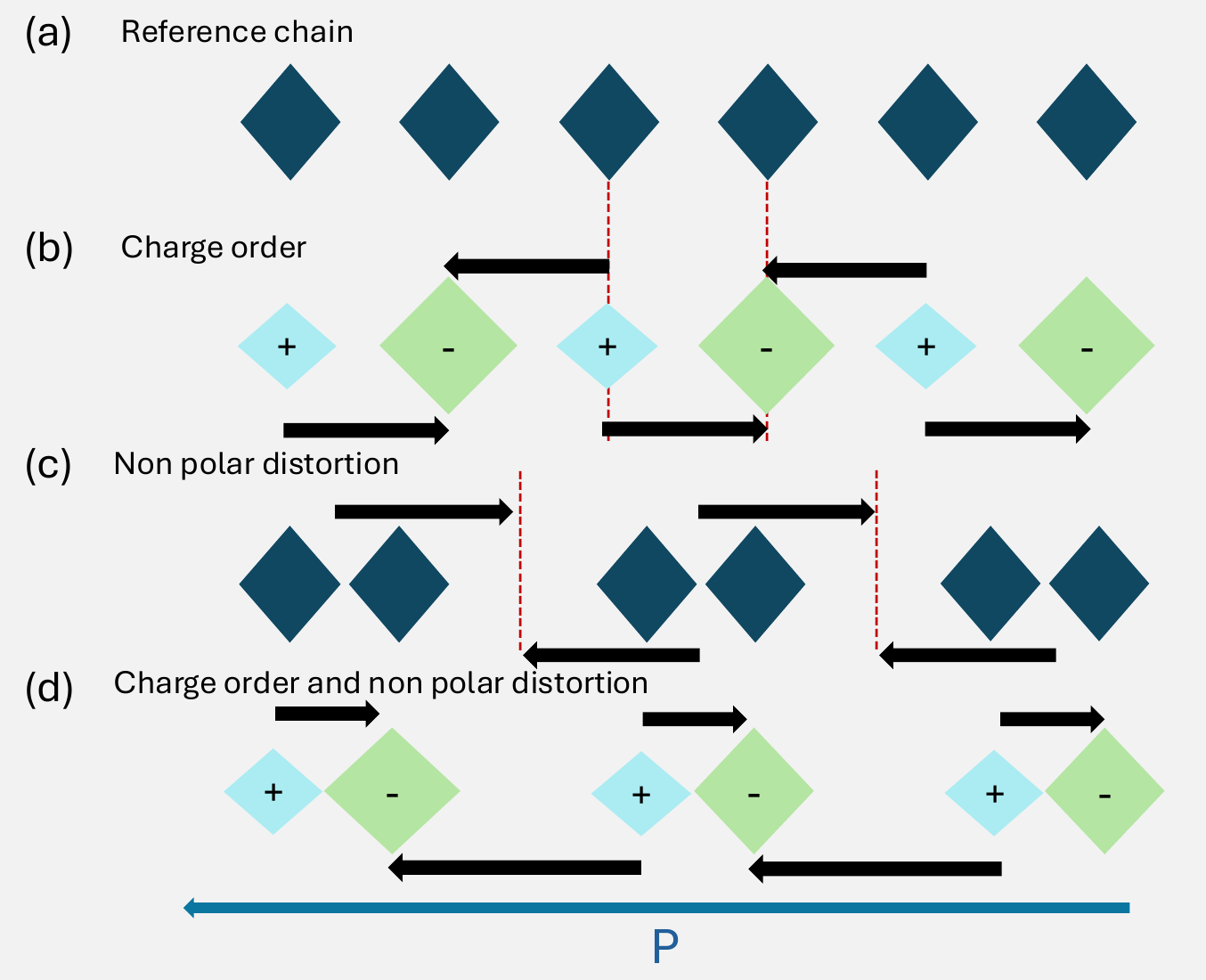}
\centering
\caption{CO-induced model of Van der Brink and Khomskii~\cite{VanDenBrink2008}. (a) Reference chain model structure with equal valence states and distances between sites. (b) CO leads to local distortion of the metal ion environment and different Wyckoff labels sets. (c) Dimerization is an example of a non-polar distortion. This distortion is in general different from the expansion and contraction of the local environments due to CO. (d) Simultaneous presence of CO and dimerization breaks inversion symmetry and originates a macroscopic polarization.}
\label{figura1}
\end{figure}

Fig.~\ref{figura1} reproduces the model of J. Van den Brink and D. I. Khomskii~\cite{VanDenBrink2008}. The reference chain structure (Fig.~\ref{figura1}(a)) displays equal valence states and equal distances between sites. CO and centrosymmetric distortions, schematically represented as alternating charge and dimerization in Figs.~\ref{figura1}(b) and~\ref{figura1}(c), respectively, maintain inversion symmetry. In both cases, electrostatic forces are symmetrical with respect to either sites or bonds, respectively, and cancel each other. However, when CO and centrosymmetric distortions are simultaneously present (Fig.~\ref{figura1}(d)), electrostatic forces are uneven, and inversion symmetry is broken. The latter induces a macroscopic polarization in the system.

Notably, CO breaks the symmetry of equivalent metal ion sites and splits the Wyckoff position labels into two or more sets, one for each valence state~\cite{Li2011}. For instance, in the reference structure of the Van den Brink-Khomskii model (Fig.~\ref{figura1}(a)), all sites belong to the same Wyckoff position label 1a. All sites are equivalent and related by translational symmetry. When CO is introduced (Fig.~\ref{figura1}(b)), the symmetry of the structure decreases and the Wyckoff positions of the reference structure split into the Wyckoff positions 1a and 1b. In each of these, the sites are equivalent, display the same valence state, and are related by a space group symmetry. 

Distinctively, CO increases and decreases the amount of localized charge at the ion sites with different valence states. This is schematically shown in Fig.~\ref{figura1}(b) by the expansion and contraction of the octahedron around the metal ion sites. In real materials, CO and Wyckoff position splitting lower the symmetry of the system and lead to additional local distortions of the ions' environments. We will refer to the structure with equal valence states and equal environment volumes as the parent structure, and to the structure with different valence states and different environment volumes as the nonpolar structure. The parent structure is, in general, a hypothetical structure, not necessarily accessible experimentally. 

We emphasize that the local distortions around the ions are not necessarily related to the $\mathbf{q} \neq 0$ zone-boundary distortion of the Van den Brink-Khomskii model (e.g. centrosymmetric dimerization in Fig.~\ref{figura1}(c)). As shown in this work, CO ferroelectrics may display centrosymmetric distortions that may or may not be associated with the CO distortion. CO-induced ferroelectrics  correspond to the cases where CO originates or couples to the polar distortion. In non CO-induced ferroelectrics, CO and polar distortions are independent and coexist in the polar phase. In both cases, secondary centrosymmetric distortions may or may not be present due to the reduction of symmetry, the presence of CO, and higher-order coupling terms. 

Despite considerable progress in the fundamental understanding of CO ferroelectrics, as shown below, the list of known CO-induced ferroelectrics remains relatively limited. The latter constrains the fundamental understanding of the phenomena, the relation between CO and lattice distortion~\cite{Balachandran2013}, and in particular, the prediction and optimization of magnetoelectric materials for real electronic and spintronic applications. In this context, accelerated discovery of materials by high throughput screening provides a way to tackle this problem. As shown below, new materials may provide relevant insight into new CO-induced ferroelectric mechanisms. 

In this work, we propose a combination of high throughput and first principles methodologies to search for new CO ferroelectrics in the Materials Project database~\cite{Jain2013}.  Here, we focus on CO-induced ferroelectrics, where CO and polar distortions are coupled and therefore display a magnetoelectric effect. We propose a set of descriptors, symmetry arguments, and first principles approaches that can be used to screen for CO-induced ferroelectrics in any general database of magnetic materials. 

Crucially, since CO is associated with a different number of electrons at otherwise equivalent sites, the cases where CO and polar distortions are effectively coupled can be identified by artificially suppressing the spin degree of freedom in the first principles simulation. The motivation behind this idea is as follows. For the particular case of magnetic systems, different valence states imply different spin values. If the spin variable is artificially suppressed from the DFT calculation, metal ions will acquire the same charge and valence state. Therefore, the polyhedron volumes around the sites will acquire the same volume, and the CO distortion will be suppressed. Therefore, if the relaxed spinless polar structure converges to a centrosymmetric structure, there must be a coupling mechanism between CO and polarization, and the ferroelectric corresponds to a CO-induced candidate. If the  polar structure remains polar after supressing the spin degree of freedom, we conclude that the polar mode is unstable by itself and, in particular, is decoupled from CO. 

Our first principles methodology applies to CO-induced ferroelectrics arising from structural phase transitions, and allows us to search for new candidates within a materials database. For these CO-induced ferroelectrics, the relation between local CO distortion and inversion symmetry breaking can be studied using symmetry-adapted modes from a hypothetical parent structure. The parent structure allows us to study the coupling between the distortion associated with CO, typically an expansion and contraction of the local environment around the metal ion, and the polar distortion present in the ferroelectric phase.

\section{Computational details}

Density functional theory (DFT) calculations are performed using the Vienna Ab initio Simulation Package (\texttt{VASP}) code~\cite{Kresse1996}. We use the generalized gradient approximation of Perdew, Burke, Ernzerhof (PBE)~\cite{Perdew1996} plus the Hubbard U correction to describe transition metal ions~\cite{Dudarev1998}. Our DFT calculations use an energy cut-off of 520~eV and symmetry adapted $\mathbf{k}$-point grids (6 $\mathbf{k}$-points per 2$\pi$ 0.25 \AA$^{-1}$ in each direction). Band gaps and Born effective charges are computed using dense grids (12 $\mathbf{k}$-points per 2$\pi$ 0.25 \AA$^{-1}$). 

Full structural relaxations of lattice parameters and atomic positions are performed until forces are smaller than 0.001 $eV/\AA$. The macroscopic polarization is estimated using the linear approximation $P \simeq e/V \sum_i Z_i \delta u_i$, where $Z_i$ and $\delta u_i$ are the Born effective charges and displacement of atom $i$, $e$ is the electron charge, $V$ is the volume of the polar structure, and the sum runs over all the ions in the structure~\cite{Resta1994}. Reported Born effective charges are obtained as $Z = Tr( Z_{\alpha \beta})/3$, with $Z_{\alpha \beta}$ the Born effective charge tensor and $Tr(\cdot)$ the trace. 

Transition metal ions display open $d$ shell orbitals; therefore, we perform collinear magnetic calculations initialized in a ferromagnetic state. The Hubbard U parameters used in this work were obtained by fitting first principles formation energy values to experimental measurements~\cite{Wang2006a, Jain2011a}. Heavy ions with open shell $f$ electrons are computationally challenging and are therefore left for future work. In addition, we neglect the effect of spin-orbit interaction and spin canting effects. 

High throughput screening is performed in the Materials Project database~\cite{Jain2013}. The Materials Project contains first-principles results for a large set of crystal structures. The majority of the materials are computed by taking experimental structures reported by the Inorganic Crystallographic Database (ICSD)~\cite{Zagorac2019} as input parameters. The Materials Project also includes hypothetical materials explored by first principles or predicted by data-driven methods~\cite{Ye2018}. 

The initial identification of ferroelectric symmetry pairs is performed with \texttt{Pyxtal}~\cite{Fredericks2021}. We use the \texttt{get$\_$transition} function to identify materials with centrosymmetric and polar structures connected by a continuous symmetry breaking, i.e. the Wyckoff positions of the centrosymmetric structure split into the Wyckoff positions of the polar structure. Then, CO candidates are selected using the \texttt{Pymatgen} code~\cite{Jain2011b}. We use the \texttt{BVAnalyzer} and \texttt{SpacegroupSymmetry} functions of \texttt{Pymatgen} to compute valence states and Wyckoff positions, respectively. \texttt{BVAnalyzer} relies on the bond valence sum (BVS) method~\cite{OKeeffe1991}. For selected cases, we analyze the structural coupling mechanism. The parent structure is obtained using the \texttt{PSEUDO} application from the Bilbao Crystallographic Server (BCS)~\cite{Capillas2011a}. Then, we use the \texttt{ISODISTORT} and \texttt{INVARIANTS} applications from the \texttt{ISOTROPY} software suite~\cite{isotropy} to analyze mode symmetries~\cite{Campbell2006} and determine the free energy coupling terms~\cite{Hatch2003}, respectively.

\section{Results}

\subsection{Symmetry mode analysis of known ferroelectrics driven by charge ordering}

\begin{figure}[htp]
\includegraphics[width=\columnwidth]{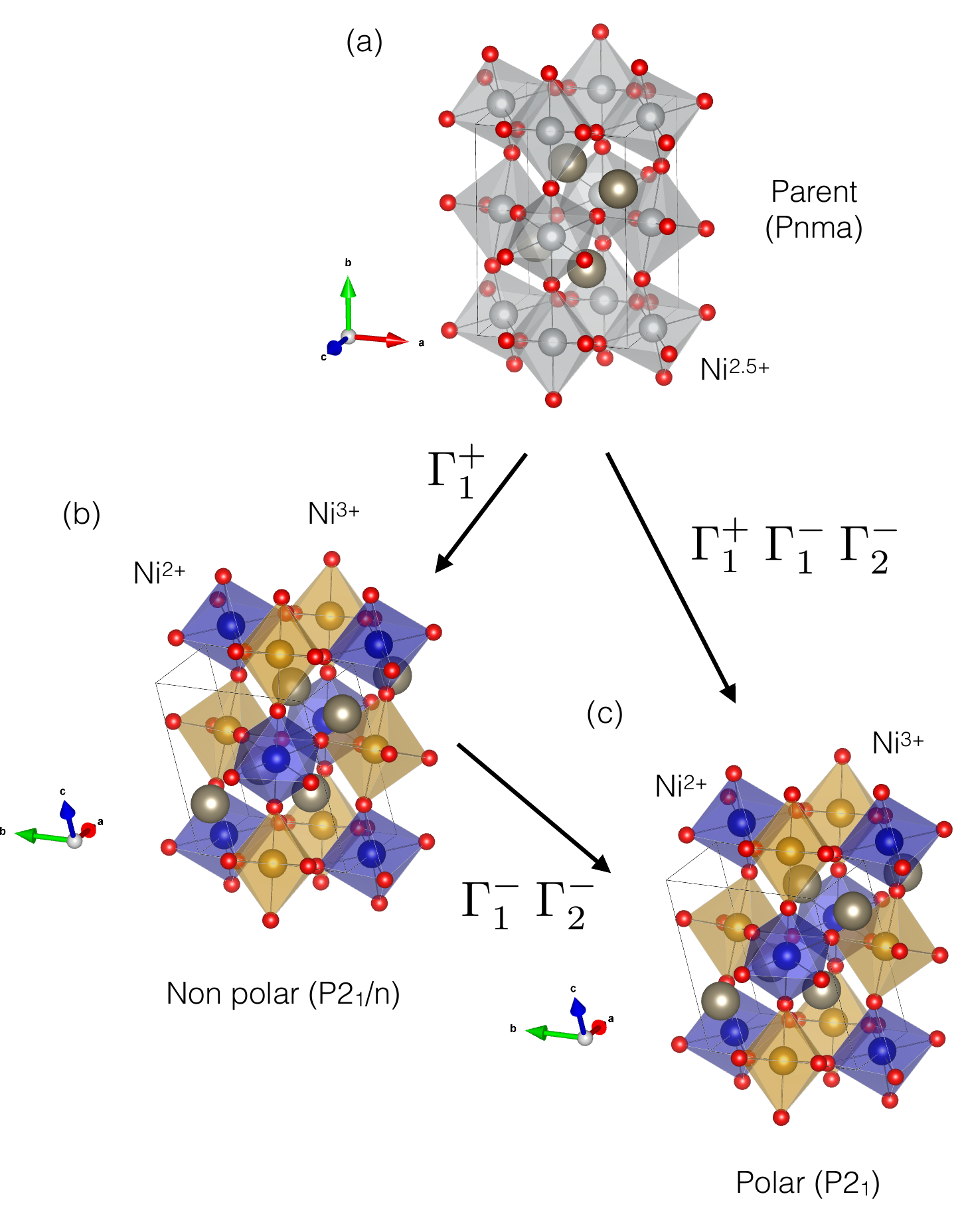}
\centering
\caption{The distortions from a hypothetical centrosymmetric \textsl{Pnma} parent structure (a) into the CO non-polar \textsl{P2$_1$/n} (b) and CO polar \textsl{P2$_1$} crystal structure (c) for TlNiO$_3$. In (a), all octahedron have the same volume and are depicted in gray color. In (b) and (c), there are two sets of octahedron volumes, larger and smaller volumes are depicted in blue and orange, respectively. Tl, Ni, and O ions are shown as dark green, gray and red spheres. The arrows show the structural modes that connect the parent structure with the non-polar and polar structures and between non-polar and polar structures.}
\label{figure2}
\end{figure}

The rare-earth nickelates \text{RNiO$_3$} with R: Ho, Lu, Pr, Nd, Sm, Tl~\cite{Alonso2000, Giovannetti2009a, Xin2014} are a well-known family of CO-induced ferroelectrics. Several members are reported in the non-polar \textsl{P2$_1$/n} and polar \textsl{P2$_1$} structures. Fig.~\ref{figure2} shows the atomic structure for the particular case of TlNiO$_3$. The non-polar \textsl{P2$_1$/n} structure displays CO with Ni$^{2+}$ and Ni$^{3+}$ in Wyckoff positions 2b and 2c, respectively (Fig.~\ref{figure2}b). The polar \textsl{P2$_1$} structure displays Ni ions with ${2+}$ and ${3+}$ CO. In this case, due to the lowering of the symmetry, the Wyckoff positions of Ni$^{2+}$ and Ni$^{3+}$ are now 2a, and the octahedron volumes around the Ni sites are different. The non-polar \textsl{P2$_1$/n} structure displays different Wyckoff position labels, whereas the polar \textsl{P2$_1$} structure displays different valence states.

The distortions induced by the CO and polar distortions are conveniently described by a hypothetical \textsl{Pnma} parent structure (see Fig.~\ref{figure2}(a))~\cite{Kim2001}. The octahedron volumes around the Ni in the \textsl{Pnma} structure are equal. The CO \textsl{P2$_1$/n} structure (see Fig.~\ref{figure2}(b)) is obtained from \textsl{Pnma} by the freezing of a $\Gamma_1^+$ mode. The $\Gamma_1^+$ mode describes the expansion and contraction of the polyhedron around the Ni sites. The polar \textsl{P2$_1$} structure is obtained from \textsl{Pnma} by the freezing of the CO $\Gamma_1^+$ and polar $\Gamma_1^-$, $\Gamma_2^-$ modes. Symmetry analysis shows third-order coupling terms between the $\Gamma_1^+$ non-polar mode and the $\Gamma_1^-$ and $\Gamma_2^-$ polar modes, of the form Q$_{\Gamma_1^+}$Q$^2_{\Gamma_1^-}$ and Q$_{\Gamma_1^+}$Q$^2_{\Gamma_2^-}$. Therefore, we conclude that, in the case of the nickelates, ferroelectricity is driven by the CO distortion~\cite{VanDenBrink2008}.

\begin{figure}[htp]
\includegraphics[width=\columnwidth]{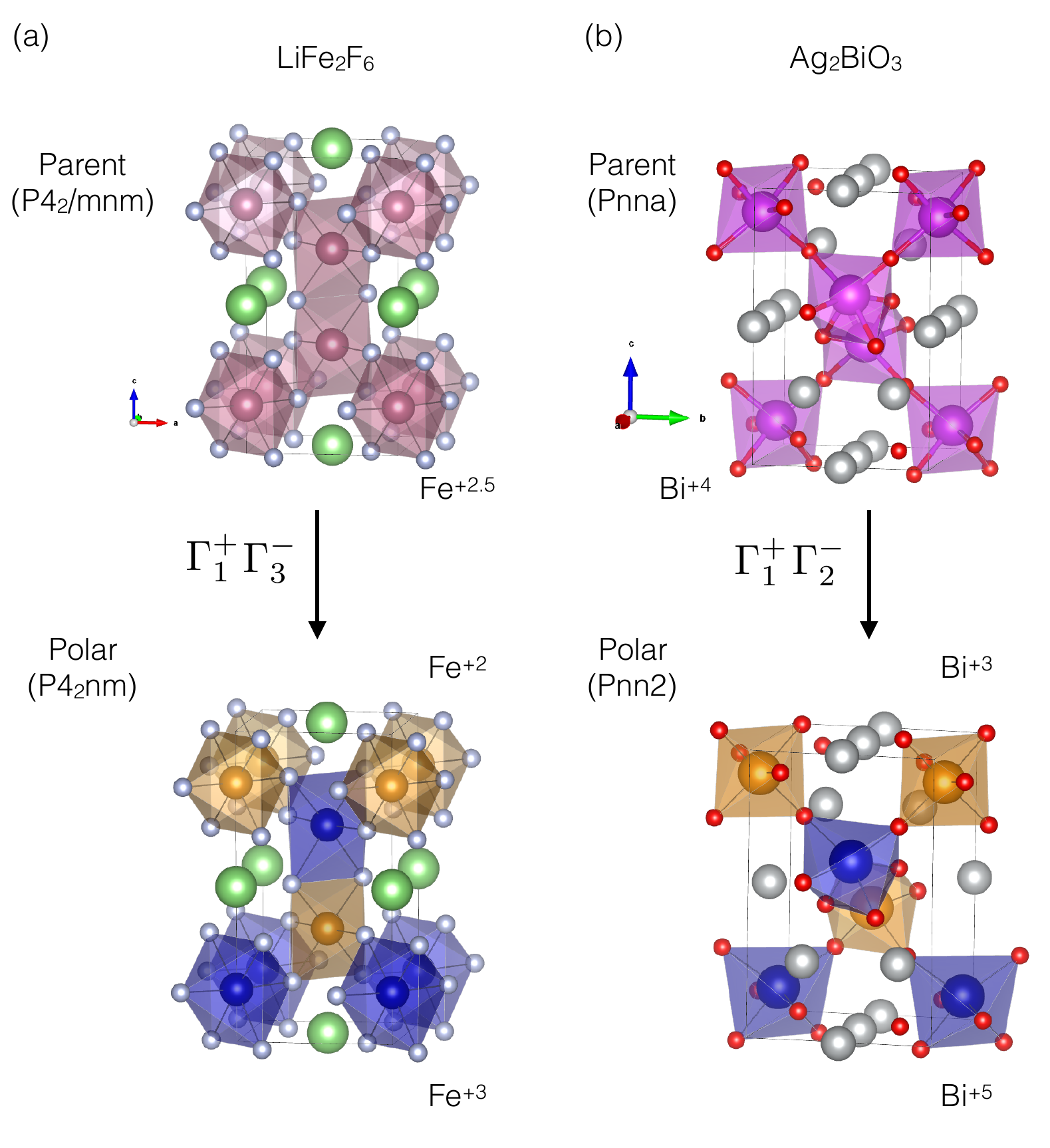}
\centering
\caption{Examples of the inversion-symmetry breaking by CO without local distortion. Parent and polar crystal structures for (a) LiFe$_2$F$_6$ and (b) Ag$_2$BiO$_3$. In the parent structures, all octahedron have the same volume, same valence state and are therefore depicted in the same color. In the polar structures, there are two sets of octahedron volumes, larger and smaller volumes are depicted in orange and blue, respectively. Li, Ag, F, and O ions are shown as green, gray, white, and red spheres, respectively. The arrows represent the structural distortion from the parent to the polar structure.}
\label{figure3}
\end{figure}

Interestingly, there are cases where the CO distortion (see Fig.~\ref{figura1}(b)) readily breaks inversion symmetry without local distortions. Fig.~\ref{figure3} shows two such examples. The CO ferroelectric LiFe$_2$F$_6$ features a polar \textsl{P$4_2$nm} structure with Fe$^{2+}$ and Fe$^{3+}$ CO~\cite{Lin2017, Fourquet1988, Greenwood1971}. The centrosymmetric \textsl{P$4_2/$mnm} parent structure displays Fe$^{2.5+}$ sites with equal volume octahedra and equivalent 4e Wyckoff positions. If the Fe$^{2+}$ and Fe$^{3+}$ CO is imposed, without further distortions, inversion symmetry is broken and LiFe$_2$F$_6$ assumes a polar \textsl{P$4_2$nm} space group symmetry. In other words, there is no CO non-polar structure available for LiFe$_2$F$_6$. The polar \textsl{P$4_2$nm} structure is obtained from the centrosymmetric parent \textsl{P$4_2$/mnm} through the combination of a $\Gamma_1^+$ nonpolar mode and a $\Gamma_3^-$ polar mode through a Q$_{\Gamma_1^+}$Q$^2_{\Gamma_3^-}$ coupling term. 

Similarly, the CO ferroelectric Ag$_2$BiO$_3$ has been reported in the centrosymmetric \textsl{Pnna} and polar \textsl{Pnn2} structures~\cite{Oberndorfer2006, He2018}. Ag$_2$BiO$_3$ displays an alternative \textsl{Pc} polar structure with both corner-sharing and non-corner-sharing octahedra, which is unrelated to \textsl{Pnna}. \textsl{Pnna} displays Bi sites with equal octahedron volumes and equivalent Wyckoff positions. However, the Bi$^{3+}$ and Bi$^{5+}$ charge disproportionation induces a phase transition to the \textsl{Pnn2} polar structure~\cite{He2018}. The polar \textsl{Pnn2} structure can be obtained from the parent \textsl{Pnna} structure by a combination of a $\Gamma_1^+$ non-polar mode and a $\Gamma_2^-$ polar mode through a Q$_{\Gamma_1^+}$Q$^2_{\Gamma_2^-}$ coupling term. 

There are other types of CO ferroelectricity that cannot be necessarily described by symmetry modes from a parent structure. These include CO ferroelectrics where inversion symmetry breaking is driven by order-disorder transitions (e.g. Fe$_3$O$_4$~\cite{Alexe2009, Yamauchi2009, Senn2012}), alloying (e.g. \text{La$_{0.5}$Ca$_{0.5}$MnO$_3$}~\cite{Radaelli1997, Efremov2004, Giovannetti2009}, Pr$_{0.5}$Ca$_{0.5}$MnO$_3$~\cite{Efremov2004, Serrao2007, Yamauchi2013}, SmBaMnO$_6$~\cite{Sagayama2014, Morikawa2012}), alloying and octahedral rotations (e.g. La$_2$NiMnO$_6$~\cite{Tang2010}, Y$_2$NiMnO$_6$~\cite{Tang2011}), atomic species occupation and octahedral rotations (Bi$_{1.5}$Sr$_{0.5}$FeCrO$_6$~\cite{Rout2018}), atomic species layering and octahedral rotations (e.g. LaFeO$_3$/LaTiO$_3$~\cite{He2016a}, LaVO$_3$/SrVO$_3$~\cite{Park2017}, LaSr$_2$V$_2$O$_7$~\cite{Park2022}) and spin canting (e.g. BiMn$_2$O$_5$~\cite{Li2011}, TbMn$_2$O$_5$~\cite{Chang2011,Wang2007,Okamoto2007}, LuFe$_2$O$_4$~\cite{deGroot2012, Ikeda2005, Angst2013}). We note in passing that CO ferroelectrics have also been engineered through layering~\cite{Balachandran2018} or tuning with an external parameter such as epitaxial strain~\cite{Rout2018}. In a few cases, CO-induced ferroelectricity is still debated~\cite{Tang2010, Lin2009, Tang2011, Serrao2007, Angst2013}. 

Therefore, having a database of materials, we first identify ferroelectric candidates by selecting materials possessing an insulating polar structure, connected to a centrosymmetric structure by a continuous symmetry distortion. Then, in close analogy with the case of the nickelates and LiFe$_2$Fe$_6$/Ag$_2$BiO$_3$, CO ferroelectrics are readily identified by their polar structures with mixed valence states, and a centrosymmetric phase which shows either: (case I) different Wyckoff position labels, or (case II) equal Wyckoff labels. The centrosymmetric structure will correspond to the non-polar structure in case I and to the parent structure in case II.

\subsection{Initial screening of ferroelectric coexisting with charge ordering}

We start by developing a set of descriptors to screen for CO ferroelectric candidates. At the time of this work, the Materials Project contained approximately 150,000 materials. From these, 29,830 were polar, and 19,107 were polar and insulating. For each of the 19,107 polar insulating structures, we search for centrosymmetric structures reported in the Materials Project having the same stoichiometry as the polar structure. Here, we do not exclude the possibility of metallic centrosymmetric structures, since CO ferroelectric switching may imply delocalization of charge between sites and hence a metallic phase~\cite{Qi2022b}. Several CO ferroelectrics display metallic centrosymmetric phases at high temperatures~\cite{Rao1998}. We focus on insulating polar materials and therefore exclude the possibility of polar metals.

Next, we use \texttt{Pyxtal} to select cases where centrosymmetric and polar structures are related by a continuous symmetry breaking of the Wyckoff positions~\cite{Zhu2022}. The latter determines a second order transition from a high symmetry centrosymmetric phase to a low symmetry ferroelectric phase. Ferroelectric materials are characterized by a small structural distortion between centrosymmetric to polar phases~\cite{Abrahams1989}. The latter condition has been previously used to identify new ferroelectrics~\cite{Smidt2020, Ricci2024}. Our final list of materials displaying centrosymmetric and polar structures related by symmetry contained 945 candidates. Note that the list contains materials with more than one centrosymmetric to polar transition, with either equal centrosymmetric or polar structures. We will denote as a ferroelectric symmetry pair the combination of a specific stoichiometry, together with a specific pair of centrosymmetric and polar structures connected by a continuous symmetry breaking. The latter does not exclude cases of ferroelectrics with near-zero polarization. However, these cases may lead to isostructural ferroelectrics with larger polarization values~\cite{Forero-Correa2025}.

\begin{figure}[htp]
\includegraphics[width=\columnwidth]{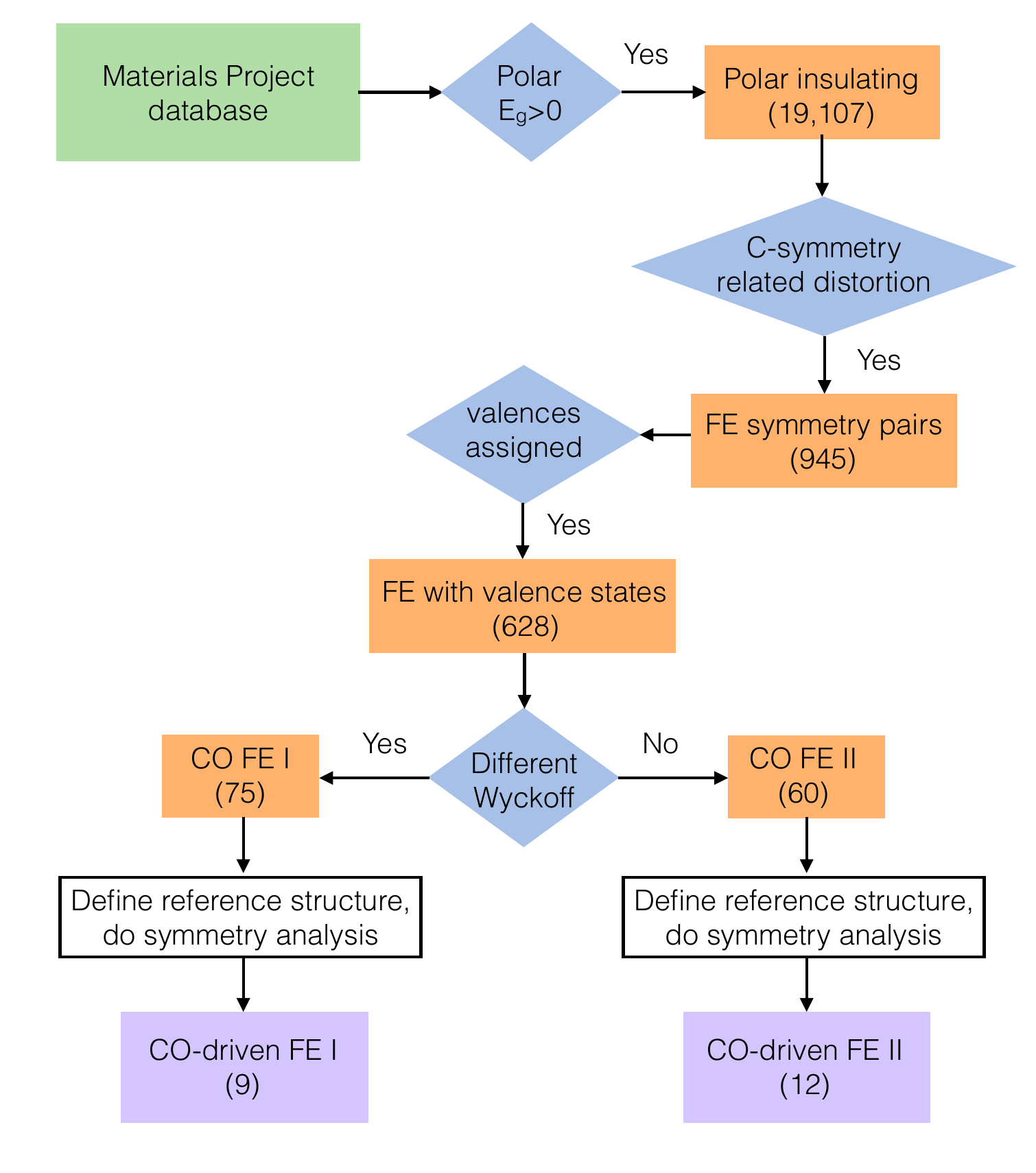}
\centering
\caption{Workflow to identify charge order (CO) ferroelectrics (FEs) with the relevant steps of the work and the number of materials considered in each step in parenthesis. The lines represent the conditions imposed between boxes. C-symmetry stands for centrosymmetric related to polar structure by a continuous symmetry distortion.}
\label{diagram}
\end{figure}

Following the analysis above, we proceed to identify polar structures with transition and post-transition metal ions in different valence states. From the initial list of 945 ferroelectric symmetry pairs, only 628 polar structures were assigned a valence state by the \texttt{get$\_$valence} function from \texttt{Pymatgen}, as shown in Fig.~\ref{diagram}. Note that one material may have more than one metal ion, and therefore, more than one CO possibility. In some cases, valence states are not assigned because the experimental parameters for the BVS method are not available. These include materials with ions such as Mo, W, Tc and several $f$ electron ions. The remaining 446 materials with unassigned valence states are left for future work.

Within the list of 628 ferroelectric symmetry pairs with assigned valence states, we identify CO ferroelectrics as those possessing different valence bands in the polar structure, and different (case I) or equal (case II) valence states in the centrosymmetric structure  (see Fig.~\ref{diagram}). The centrosymmetric structure will correspond to the non-polar and parent structures in case I and II, respectively. Tables S1 and S2 in the Supplementary Material in Ref.~\cite{supp} show the resulting 80 and 67 candidate CO ferroelectrics for cases I and II, respectively, along with relevant information extracted from the Materials Project. Note that some cases have equal stoichiometry but different symmetry pairs (e.g. Fe$_3$O$_4$), because either the centrosymmetric, polar, or both structures are different. Cases with equal stoichiometry are labeled with a number after the stoichiometry (e.g. Fe$_3$O$_4$-1, Fe$_3$O$_4$-2, etc.). Tables S1 and S2 include Materials Project identification numbers, space group symmetries, and band gaps for centrosymmetric and polar structures, as well as energy and volume differences between centrosymmetric and polar structures, and the energy above hull for the polar structure from the Materials Project. 

Tables S1 and S2 show 44 and 48 cases, respectively, where the polar ferroelectric phase is the ground state ($\Delta E<0$). In case I, materials where the non-polar structure is the ground state, and the nonpolar-polar energy difference is small, may be considered as antiferroelectric candidates~\cite{Reyes-Lillo2014}. However, further calculations are required to explore the energy landscape between the structures. The energy difference between non-polar and polar structures may be controlled with external effects such as pressure, strain, or chemical composition. Note that several non-polar and parent structures are metallic and are therefore unsuitable as a reference for Berry phase polarization calculations. We later estimate the polarization of the final candidates using Born effective charges. The energy above hull can be used to estimate the likelihood of the material to be synthesized; values~$<0.01$ eV/atom are expected to be accessible to synthesis.  

The CO ferroelectric candidates listed in Tables S1 and S2 display a coexistence of polar insulating structure and CO. However, CO and polar distortions are not necessarily related. In addition, some of the materials display a relatively large energy difference ($>500$~meV per atom) between the polar and nonpolar structures compared to traditional ferroelectrics, where energy differences are typically below 200~meV per atom~\cite{Ricci2024}. The latter suggests that many of these systems may not be switchable with an external electric field. In addition, few candidates display relatively unphysical volume expansion ($>5$\%), which may arise due to mathematically feasible distortions among radically different structures. To the best of our knowledge, with the exception of LiFe$_2$F$_6$ and Fe$_3$O$_4$, none of these materials has been previously proposed as CO ferroelectric.

We further note that CO ferroelectrics without a coupling between CO and ferroelectricity may be used to construct CO-induced heterostructures, order-disorder, or layered systems. Several materials included in Tables S1 and S2 may display other functional or electronic properties of potential interest. In addition, we note that a coupling mechanism may arise when spin-orbit interaction and non-collinear magnetic effects are included in the system. We leave this for future work.

\subsection{Filtering ferroelectrics driven by charge ordering}

Here, we use the intuition acquired in Section A from known CO-induced ferroelectrics to search for new candidates among the initial screening of Section B. Tables S3 and S4 report the magnetic ground state configuration for the 55 and 64 distinct polar structures reported in Tables S1 and S2, respectively~\cite{Horton2019}. Interestingly, the great majority of these polar structures, 48 of 55 (87$\%$) and 53 of 64 (83$\%$) in Tables S3 and S4, respectively, are reported as magnetic in the Materials Project, with nonzero local spin in the metal ions. In total, considering cases I and II, there are only 18 non-magnetic cases.

\begin{figure}[htp]
\includegraphics[width=\columnwidth]{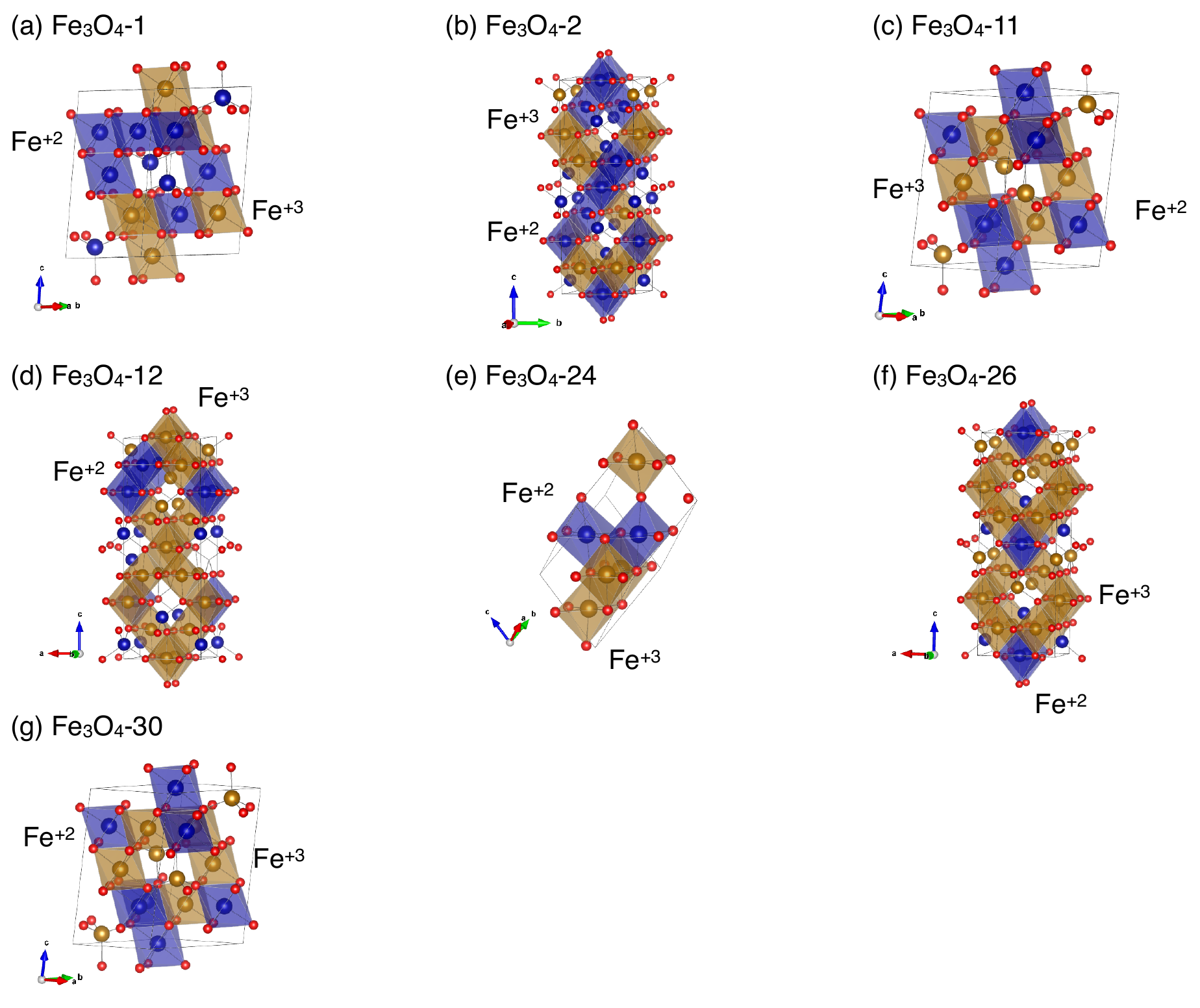}
\centering
\caption{Polar charge-ordered structures for Fe$_3$O$_4$. Only the octahedral coordination polyhedron around the charge-ordered ion is considered; other coordination polyhedra (e.g., tetrahedra) are excluded. The different colors denote the CO displayed by the ions octahedra. The numbers next to the formula distinguish between ferroelectric symmetry pairs with equal polar structure but different centrosymmetric structure.}
\label{grupo0}
\end{figure}

In the following, we discard the 18 non-magnetic cases and focus on the ferromagnetic spin configuration of the 101 remaining materials. This implies that all selected CO-induced ferroelectrics in this study are intrinsically multiferroic. However, we note that the ferromagnetic configuration may not be the lowest magnetic state of the system. As motivated earlier, eliminating the non-magnetic material will allow us to identify the cases where CO and polarization are effectively coupled. For the 48 and 53 resulting materials in Tables S3 and S4, we perform the computational experiment introduced earlier and consider the stability of the CO distortion in the polar structure when the spin variable is artificially suppressed from the DFT simulation.

Tables S3 and S4 show the space groups before and after removing the spin degree of freedom from the DFT calculation, and full relaxation of lattice parameters and internal atomic positions.  We find that 18 and 18 polar structures from Tables S3 and S4, respectively, converge to a structure with a centrosymmetric space group and therefore are candidates for CO-induced ferroelectrics (highlighted in green). In these cases, the symmetry of the structure increases, and the sites with different sets of Wyckoff positions in the polar structure are now described by the same Wyckoff position label in the relaxed structure. For 19 and 26 materials in case I and II, respectively, the initial polar structure remains in the same polar space group, whereas for 11 and 9 materials in case I and II, respectively, the initial polar structure transforms to a higher-symmetry polar space group. In both situations, the local distortions around the sites become symmetrized; however, in the first case, the symmetrization of the octahedron environments around the sites is not enough to change the global space group symmetry.

For simplicity, we focus on materials displaying CO within an octahedral environment. Therefore, we disregard Li$_6$(CoO$_2$)$_5$ and LiMn$_2$(BO$_3$)$_2$ with coordination 4 and 5 for Co and Mn, respectively. Similarly, we disregard 4 cases with a single valence state within the octahedral environments. For instance, Y$_{4}$Fe$_{13}$Si$_{2}$(SbO$_{14}$)$_{2}$ displays Fe$^{3+}$ and Fe$^{4+}$ within octahedron and tetrahedron local environments, respectively. Therefore, Fe adopts only the $3+$ valence states within the octahedron, and the charge is ordered within the octahedral local environment. 

Notably, the list of materials that transform from polar to a centrosymmetric structure in the absence of spin (highlighted in green in Tables S3 and S4) includes LiFe$_2$F$_6$ (see Fig.~\ref{figure2}) and 8 entries of the well-known CO-induced ferroelectric ferrimagnet Fe$_3$O$_4$~\cite{Alexe2009, Yamauchi2009, Senn2012}. Fe$_3$O$_4$ crystallizes in the \textsl{Fd$\bar{3}$m} inverted cubic spinel structure at room temperature. The tetrahedral sites are occupied by Fe$^{3+}$ cations, whereas the octahedral sites are occupied by an equal number of randomly distributed Fe$^{2+}$ and Fe$^{3+}$ cations~\cite{Wang2014}. Below 120~K, Fe$_3$O$_4$ undergoes a sharp first-order transition, and the structure assumes a CO due to the random distribution of Fe$^{2+}$ and Fe$^{3+}$ ions~\cite{Verwey1939}. Fe$_3$O$_4$ has been experimentally reported in space group symmetries Hematite \textsl{P1}~\cite{Verdugo-Ihl2020} and \textsl{Cc} (\textsl{Pmc2$_1$} in pseudocubic axis)~\cite{Iizumi1982}.

Fig.~\ref{grupo0} shows the CO polar structures for Fe$_3$O$_4$. Fe$^{2+}$ and Fe$^{3+}$ ions are denoted as blue and orange octahedra, respectively. These simulated structures can be interpreted as commensurate ordered snapshots of the experimentally disordered structure. An accurate description of the CO ground state of Fe$_3$O$_4$ would require simulating an incommensurate and disordered distribution of Fe$^{2+}$ and Fe$^{3+}$ ions. Table S5 reports our computed lattice parameters, total energy, and volume changes after full structural relaxations for the different structures of Fe$_3$O$_4$ identified by our work. We find that the ground state structure has \textsl{Pbcm} symmetry. The lowest energy polar structure is \textsl{Pmc2$_1$} with 3.38 meV above the ground state. 

Table S6 reports the Wyckoff positions, BVS states, Born effective charges, average bond lengths, and octahedral volumes for the centrosymmetric and polar structures along the CO ferroelectric transition of Fe$_3$O$_4$. Valence states computed with the BVS method and Born effective charges computed with DFT are well correlated with octahedron volumes. Careful inspection of Table S6 shows that the valence states of the Fe ions change from the centrosymmetric to the polar structure. This behavior is expected for CO-induced ferroelectrics, where CO and polar distortion are coupled.


\begin{table}
\centering
\caption{CO-induced ferroelectrics obtained as the final result of our high-throughput screening. Materials classified as case I and case II ferroelectrics are shown separately (see Fig.~\ref{diagram}). For each material we report the parent, non-polar and polar space group symmetries (R-SG, NP-SG, P-SG) and band gaps (E$_g^{R}$, E$_g^{NP}$, E$_g^{P}$; eV), as well as the averaged local low (LS) and high (HS) spin ($\mu_B/2$) for CO ions with mixed valence states in the polar structure. Materials are reported in alphabetical order.}
\begin{tabular}{llll rrrr}
\hline \hline
Case & Material             & NP-SG & P-SG     & E$_g^{NP}$ &  E$_g^{P}$ & LS & HS \\ \hline

I &Fe$_4$As$_5$O$_{13}$     & \textsl{P$\bar{1}$}  & \textsl{P1}     & 3.36 &  3.66         & 3.77  & 1.63\\
&Li$_4$Co$_5$SnO$_{12}$     & \textsl{C2/m}        & \textsl{P1}     & 0.11 &  0.34         & 1.94  & 0.59 \\
&Li$_4$TiCo$_5$O$_{12}$     & \textsl{C2/m}        & \textsl{P1}     & 0.72 &  0.76         & 1.25  & 0.28 \\ 
&Li$_5$Ti$_2$Fe$_3$O$_{10}$ & \textsl{P$\bar{1}$}  & \textsl{P1}     & 2.65 &  2.70         & 4.27  & 3.74 \\
&Li$_7$Co$_5$O$_{12}$       & \textsl{C2/m}        & \textsl{P1}     & 0.00 &  0.46         & 2.89  & 0.50 \\
&Li$_9$Co$_7$O$_{16}$       & \textsl{C2/m}        & \textsl{P1}     & 0.00 &  0.00         & 2.26  & 1.00 \\
&Li(Fe$_2$O$_3$)$_4$        & \textsl{P$\bar{1}$}  & \textsl{P1}     & 2.96 &  2.97         & 4.33  & 3.82 \\
&LiCr$_2$O$_4$              & \textsl{P4$_3$22}    & \textsl{P1}     & 0.69 &  0.79         & 2.95  & 2.19 \\
&V$_3$(O$_2$F)$_2$          & \textsl{Cmmm}        & \textsl{Amm2}   & 0.48 &  0.59         & 1.92  & 1.21 \\
\hline
Case & Material             & R-SG & P-SG     & E$_g^{R}$ &  E$_g^{P}$ & LS & HS \\ \hline

II &AlNi(WO$_4$)$_2$        &  \textsl{P$\bar{1}$}   & \textsl{P1}      &  0.52 & 1.60         & 0.89 & 0.05 \\
&Fe$_2$BO$_4$               &  \textsl{P2$_1$/c}     & \textsl{Pc}      &  0.18 & 1.63         & 3.78 & 2.18 \\
&Li$_2$Fe(CoO$_3$)$_2$      &  \textsl{C2/m}         & \textsl{C2}      &  0.00 & 0.33         & 3.28 & 0.57 \\ 
&Li$_2$MnV$_3$O$_8$         &  \textsl{P$2_12_12_1$} & \textsl{P1}      &  0.88 & 1.14         & 2.88 & 1.09 \\
&Li$_2$Mn$_3$VO$_8$         &  \textsl{P2$_13$}      & \textsl{P2$_1$}  &  0.01 & 0.75         & 3.88 & 2.15 \\
&Li$_5$Mn$_6$O$_{12}$       &  \textsl{P$\bar{1}$}   & \textsl{P1}      &  0.00 & 0.84         & 3.85 & 3.19 \\
&LiFe$_2$O$_3$              &  \textsl{P$\bar{3}$m1} & \textsl{P1}      &  0.00 & 0.61         & 4.30 & 3.73 \\
&Li(FeO$_2$)$_2$            &  \textsl{C2/c}         & \textsl{P1}      &  0.00 & 0.00         & 4.14 & 4.14 \\ 
&MnAl(WO$_4$)$_2$           &  \textsl{P2/c}         & \textsl{Pc}      &  0.86 & 1.62         & 0.94 & 0.13 \\
&Mo$_2$P$_3$O$_{13}$        &  \textsl{C2/c}         & \textsl{Cc}      &  0.00 & 1.71         & 0.26 & 0.02 \\
&NaV$_2$O$_4$               &  \textsl{Pnma}         & \textsl{Pmn2$_1$}&  0.00 & 0.04         & 1.76 & 1.26 \\
&V$_6$O$_{13}$              &  \textsl{P2$_1$/c}     & \textsl{Pc}      &  0.00 & 0.04         & 1.08 & 0.23 \\

\hline \hline
\end{tabular}
\label{table2}
\end{table}

\subsection{Analysis of ferroelectric candidates using symmetry-adapted modes}

\begin{figure}[htp]
\includegraphics[width=\columnwidth]{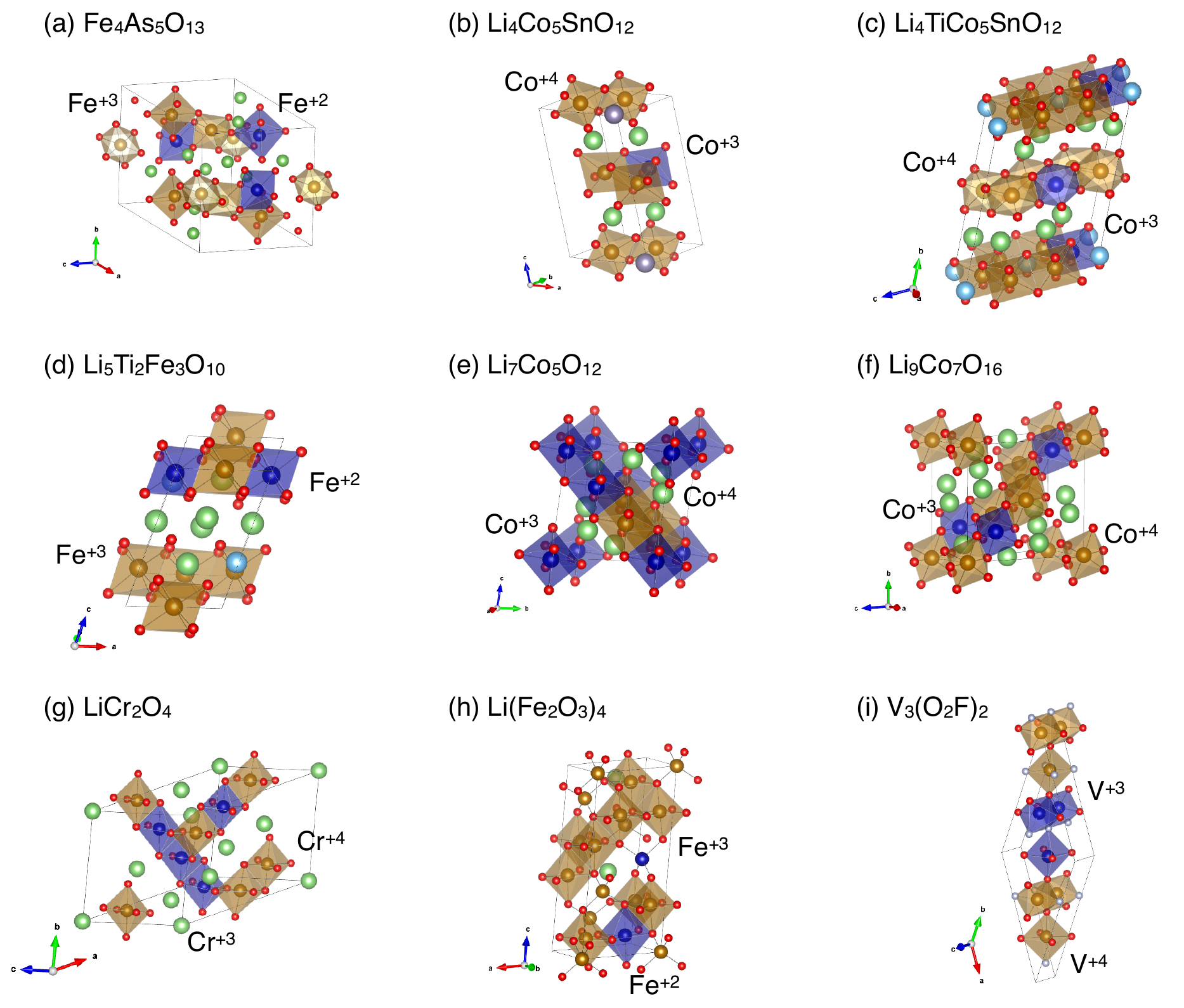}
\centering
\caption{Polar structures of CO-induced candidates for case I, see Table~\ref{table2}. The CO is visualized by the blue and orange octahedral coordination environments around the metal ions. Blue and orange octahedrons denote larger and smaller octahedron volumes, associated to localization and delocalization of charge, respectively. The valence states of the ions are shown next to each structure. Tetrahedron local environments are not shown.}
\label{grupo1}
\end{figure}

\begin{figure}[htb]
\includegraphics[width=\columnwidth]{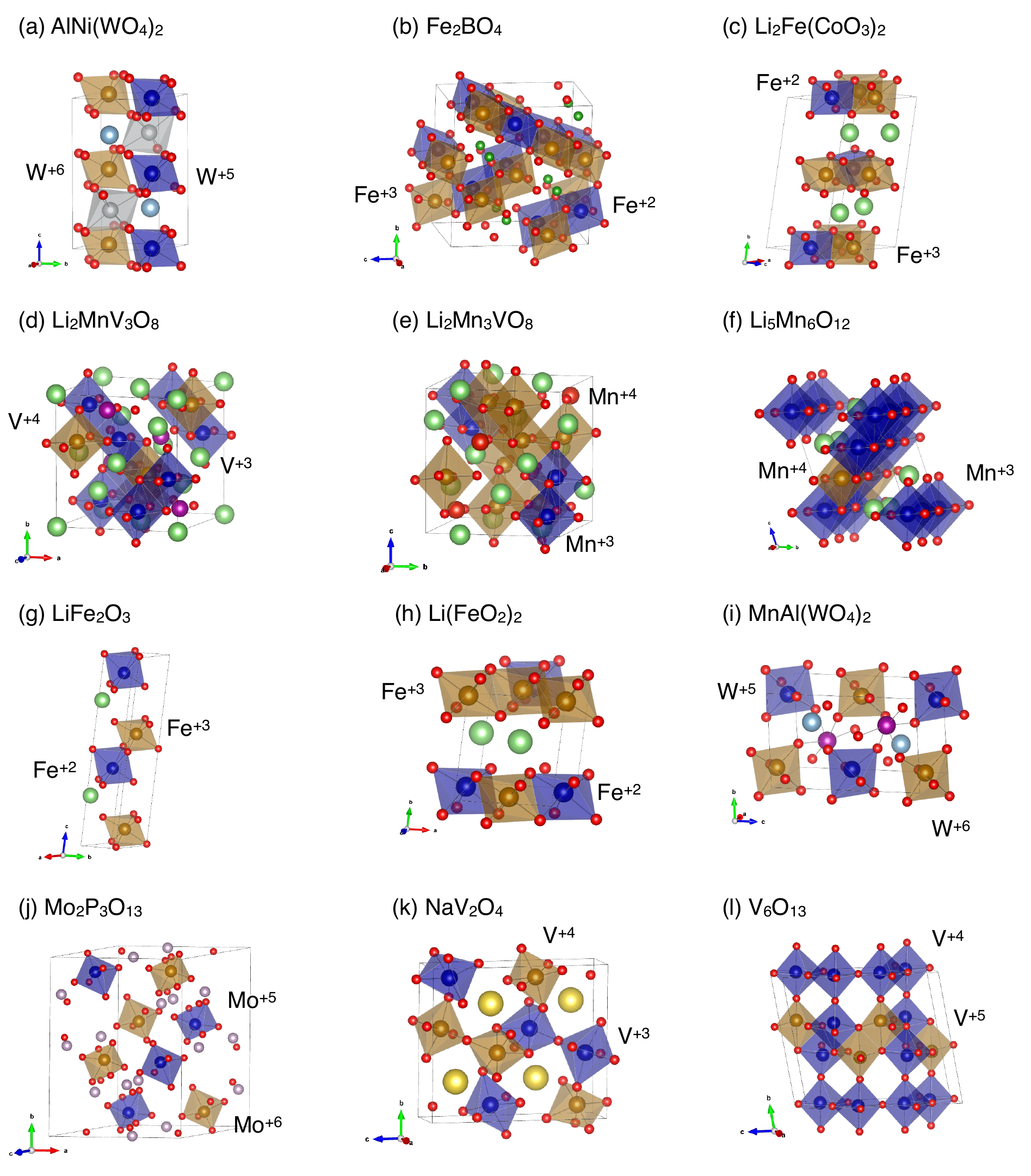}
\centering
\caption{Same as Fig.~\ref{grupo1}, but for case II CO-induced ferroelectrics.}
\label{grupo2}
\end{figure}

Now, we focus on the remaining 21 candidates, after eliminating known, non-octahedral and non-CO cases, representing the final result of our high-throughput search of CO-induced ferroelectrics. Table~\ref{table2} summarizes the 21 candidate materials, comprising 9 case I and 12 case  II CO-induced ferroelectrics. To our knowledge, these materials have not been previously proposed as CO ferroelectrics.  Table~\ref{table2} reports the space group symmetry and band gaps for non-polar and polar structures (case I) and for parent and polar structures (case II), as well as the local spin values of the polar structure. The spin values for low and high valence states are obtained for the ferromagnetic phase and averaged over equivalent sites. Similar to Fe$_3$O$_4$, there are two other materials displaying more than one non-polar to polar transition, namely Li$_4$TiCo$_5$O$_{12}$ and Li$_2$Fe(CoO$_3$)$_2$ (see Tables S1 and S2). For both cases, Table~\ref{table2} reports the ferroelectric transitions to the lowest energy polar structure, with \textsl{P1} and \textsl{C2} space group symmetry, respectively. Figs.~\ref{grupo1} and~\ref{grupo2} show the polar crystal structures of the candidates in Table~\ref{table2}. The CO of the structure is shown with blue and orange octahedra. 

Full characterization of all CO ferroelectric transitions is provided in Tables S7 and S8. These report the Wyckoff positions, BVS states, Born effective charges, average bond lengths, and octahedral volume along the CO transition from non-polar to polar (case I) and parent to polar (case II). We find a clear correlation between valence states and octahedron volumes, suggesting that valence states obtained from the BVS method are accurate enough to identify CO. Average bond lengths display subtle differences within the structure and cannot discern valence states. Previous work has used Born effective charges as a means to identify CO~\cite{Wang2014}. We find that Born effective charges display clear deviations from valence states and octahedron volumes, and therefore are rather unsuitable to screen CO. In addition, Born effective charges display different values within the same Wyckoff position label, showing a strong dependence on the local geometry. 

\begin{table*}[hbt]
\centering
\caption{Space group symmetries and distortions for the  CO-induced ferroelectrics from Table~\ref{table2}. For each material we report parent, non-polar and polar space group symmetries (R-SG, NP-SG, P-SG), CO (Q$_1$) and polar symmetry modes distortion (Q$_2$), amplitudes ($| Q_1 |$ and $| Q_2 |$), energy difference ($\Delta E$; meV/atom), volume change ($\Delta V$; \%) and polarization (P; $\mu C/cm^2$). $\Delta E = E_{P} - E_{NP}$, where $E_{NP}$ and $E_{P}$ represent the non-polar (NP) and polar (P) total energies. $\Delta E < 0$ implies polar phase is lower in energy. $\Delta V = (V_{P}- V_{NP}) \times 100/V_{NP}$, where $V_{NP}$ and $V_{P}$ are the NP and P volumes. In case II, CO readily induces the polar structure, and there is no non-polar structure. Materials are reported in alphabetical order.}

\begin{tabular}{lllll cc cc rrr}
\hline \hline
Case & Material  &  R-SG & NP-SG  &  P-SG    & Q$_1$ & Q$_2$ &  $| Q_1 |$ & $| Q_2 |$ & $\Delta E$   & $\Delta V$ & Pol. \\ \hline
I &Fe$_4$As$_5$O$_{13}$  & \textsl{P$\bar{1}$}   & \textsl{P$\bar{1}$} & \textsl{P1} & $\Gamma_1^+$ &  $\Gamma_1^-$ &  0.62 & 0.76 & 9.0  &  -1.0 &  1.6 \\
&Li$_4$Co$_5$SnO$_{12}$  & \textsl{C2/m}  & \textsl{C2/m} & \textsl{P1} & $\Gamma_1^+$ &  $\Gamma_1^-$  &  0.31 & 0.22 & -59.3  &  1.5 & 15.7 \\

&Li$_4$TiCo$_5$O$_{12}$    & \textsl{C2/m} & \textsl{C2/m} & \textsl{P1} & $\Gamma_1^+$ & $\Gamma_1^-$ $\Gamma_2^-$ &   0.25 & 0.22 0.04 & 4.9 &0.5 & 111.6\\

&Li$_5$Ti$_2$Fe$_3$O$_{10}$    & \textsl{P$\bar{1}$}   & \textsl{P$\bar{1}$} & \textsl{P1} & $\Gamma_1^+$ &  $\Gamma_1^-$ &  0.35 & 0.46 & -5.5  &  0.6 & 6.4  \\
&Li$_7$Co$_5$O$_{12}$   & \textsl{C2/m}  & \textsl{C2/m} & \textsl{P1} & $\Gamma_1^+$ &  $\Gamma_2^-$  & 0.15 & 0.22&  7.9  &  0.9& 0.3 \\
&Li$_9$Co$_7$O$_{16}$   & \textsl{C2/m}   & \textsl{C2/m} & \textsl{P1} &  $\Gamma_1^+$ &  $\Gamma_1^-$ &  0.29 & 0.25 & -19.6  &  -1.2& 72.0   \\
&Li(Fe$_2$O$_3$)$_4$    & \textsl{C2/c} & \textsl{P$\bar{1}$} & \textsl{P1} & $\Gamma_1^+$ &  $\Gamma_2^-$  & 0.29 & $<$0.05 & -4.6  &  0.3 & 1.7  \\
&LiCr$_2$O$_4$ & \textsl{I4$_1$/amd}    & \textsl{I4$_1$/amd} & \textsl{Pc}    & $\Gamma_1^+$ &  $\Gamma_5^-$  & 0.86 & 0.17 & -3.7  &  0.6 & 49.3  \\
&V$_3$(O$_2$F)$_2$      & \textsl{Cmmm}   &\textsl{Cmmm} & \textsl{Amm2}   & $\Gamma_1^+$ &  $\Gamma_4^-$ &  0.27 & 0.17& 0.5  &  0.0& 69.0  \\

\hline

II& AlNi(WO$_4$)$_2$ &   \textsl{P$\bar{1}$}  & $-$     & \textsl{P1}        & $\Gamma_1^+$& $\Gamma_1^-$  &  0.67 &  0.17 & -186.6 & 1.2 & 24.8 \\  
& Fe$_2$BO$_4$            &    \textsl{P2$_1$/c}  & $-$        & \textsl{Pc}        & $\Gamma_1^+$  & $\Gamma_2^-$ &  0.75 & 0.06&  -627.2 & 15.7 & 34.6 \\  
& Li$_2$Fe(CoO$_3$)$_2$ &     \textsl{C2/m}     & $-$            & \textsl{C2}        & $\Gamma_1^+$  & $\Gamma_1^-$  &  1.08 & 2.03 & -174.1 & 9.0 & 68.9\\ 
& Li$_2$MnV$_3$O$_8$      & \textsl{P$2_12_12_1$} &  $-$   & \textsl{P1}        & $\Gamma_1$ & $\Gamma_2$ $\Gamma_3$ $\Gamma_4$ &  1.28 & 0.11 0.12 0.12 & -330.8 & 8.5 & 0.5 \\ 

& Li$_2$Mn$_3$VO$_8$      &   \textsl{P2$_1$3}   & $-$       & \textsl{P2$_1$}    & $\Gamma_1$ & $\Gamma_4$ &  0.32 &  0.86 & -554.6 & 8.2 & 26.8 \\  
&Li$_5$Mn$_6$O$_{12}$  & \textsl{P$\bar{1}$} & $-$ & \textsl{P1}&$\Gamma_1^+$ &  $\Gamma_1^-$  & 0.40 & 1.32 & -760.7 & 11.4 & 0.3 \\  
& LiFe$_2$O$_3$           & \textsl{P$\bar{3}$m1} &  $-$     & \textsl{P1}        & $\Gamma_1^+$ &  $\Gamma_2^-$ $\Gamma_3^-$ &  0.05 & 0.56 1.68 & -702.3& 17.9 & 4.5 \\  
& Li(FeO$_2$)$_2$         & \textsl{C2/c}  &  $-$             & \textsl{P1}        & $\Gamma_1^+$ &  $\Gamma_1^-$ $\Gamma_2^-$ &  0.64 & $<$0.01 & -609.6 & 11.3 & 0.5\\ 
& MnAl(WO$_4$)$_2$        & \textsl{P2/c}  &  $-$             & \textsl{Pc}        & $\Gamma_1^+$& $\Gamma_2^-$ &  4.83 & 0.23  &-325.6  & 7.2 & 12.1\\  
& Mo$_2$P$_3$O$_{13}$     & \textsl{C2/c}  &  $-$             & \textsl{Cc}        & $\Gamma_1^+$& $\Gamma_2^-$ &  0.24 & 0.36 & -47.1 & -1.1 & 0.1 \\  
& NaV$_2$O$_4$          & \textsl{Pnma}  &  $-$        & \textsl{Pmn2$_1$}  & $\Gamma_1^+$& $\Gamma_2^-$& 0.41 & 0.21 & -281.9 & 5.0 & 8.1\\  
& V$_6$O$_{13}$           & \textsl{P2$_1$/c} & $-$            & \textsl{Pc}        & $\Gamma_1^+$ &  $\Gamma_2^-$ & 0.35 & 0.21 & -107.9 & 2.5 & 1.4\\  
\hline \hline  
\end{tabular}
\label{table4}
\end{table*}

The parent structures of the 12 materials in Table S8 can be separated into two groups. In 4 cases, namely Li$_5$Mn$_6$O$_{12}$, LiFe$_2$O$_3$, Li(FeO$_2$)$_2$, and Mo$_2$P$_3$O$_{13}$, the metal ions in the parent structure belong to a single Wyckoff position label set. In these cases, the \texttt{Pymatgen} code is not able to assign valence states. From symmetry considerations, and in complete analogy to LiFe$_2$F$_6$ and Ag$_2$BiO$_3$, the valence states of the metal ions in the parent structures are expected to be half-integers. For the other 8 cases, the parent structure features more than one ionic group with the same Wyckoff position label. In all 8 cases, there is one specie with two or more sets of Wyckoff position labels. For example, V in V$_6$O$_{13}$ appears in three different 4e sets. In these 8 cases, \texttt{Pymatgen} is able to assign the valence states in the parent structure. In 5 cases, namely AlNi(WO$_4$)$_2$, Li$_2$Fe(CoO$_3$)$_2$, Li$_2$MnV$_3$O$_8$, Li$_2$Mn$_3$VO$_8$, and MnAl(WO$_4$)$_2$, there are two different ion species with the same Wyckoff position label. These systems correspond to materials with two or more simultaneous COs in the parent structure, one for each set of Wyckoff position label set.

Note that in all observed cases, except Li$_4$Co$_5$SnO$_{12}$ and V$_6$O$_{13}$, the valence states of the metal ions change from the non-polar to the polar structures. As the polar distortion freezes in, the resulting polyhedral distortions change the valence states of the CO ions. This is indeed expected for a CO-induced ferroelectric, where polar and CO distortions are coupled. 

Next, we explore the structural coupling mechanisms for the candidate CO ferroelectric materials reported in Table~\ref{table2}. To this end, we identify a hypothetical parent structure for both the non-polar and polar structures. The parent structure allows us to (i) decompose the centrosymmetric and polar distortions using symmetry-adapted modes, (ii) identify the specific modes associated with the CO in the nonpolar structure and inversion symmetry breaking in the polar structure, and (iii) identify the coupling term between CO and polar modes in the Landau free energy expansion driving the ferroelectric transition.

For 7 systems in case I, namely Fe$_4$As$_5$O$_{13}$, Li$_4$Co$_5$SnO$_{12}$, Li$_4$TiCo$_5$O$_{12}$, Li$_5$Ti$_2$Fe$_3$O$_{10}$, Li$_7$Co$_5$O$_{12}$, Li$_9$Co$_7$O$_{16}$, and Li(Fe$_2$O$_3$)$_4$, and for all systems in case II, the centrosymmetric structure obtained from the spinless structural relaxations is readily a parent symmetry structure. In 2 cases, LiCr$_2$O$_4$ and V$_3$(O$_2$F)$_2$, the resulting structure is not a supergroup of the original non-polar phase.  For these, we employed the \texttt{PSEUDO} function in BCS to find a parent structure.  

\begin{figure}[htp]
\includegraphics[width=\columnwidth]{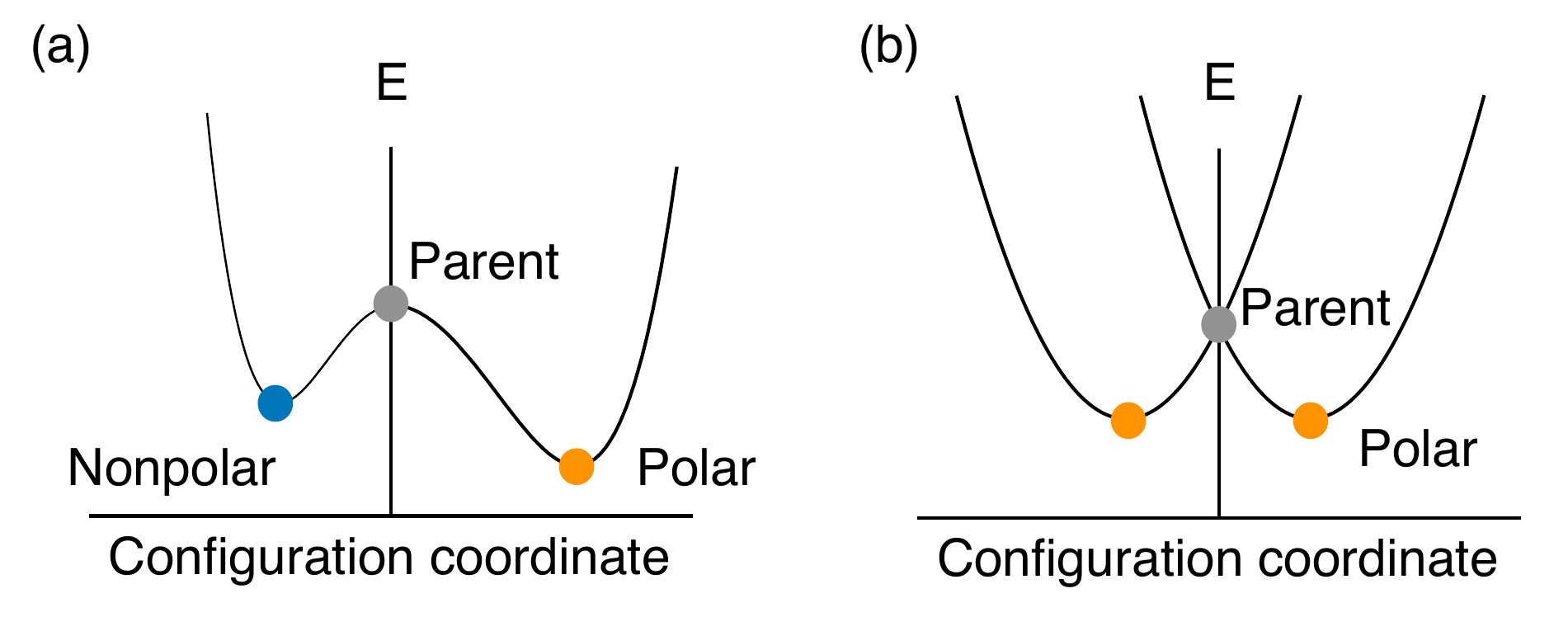}
\centering
\caption{Schematic energy (E) versus configuration coordination profiles for (a) case I and (b) case II CO ferroelectrics. In case I, non-polar and polar structures can be obtained from a parent structure. In case II, CO breaks inversion symmetry, and the polar structure is readily obtained from the parent structure and there is no non-polar structure.}
\label{figure4}
\end{figure}

Table~\ref{table4} reports the space group symmetry for parent, non-polar, and polar structures, symmetry and amplitudes of the symmetry modes connecting the parent structure with non-polar and polar phases, as well as the energy difference, volume change between non-polar to polar (case I) and parent to polar (case II), and ferroelectric polarization of the CO-induced ferroelectrics from Table~\ref{table2}. Fig.~\ref{figure4} shows a schematic of the energy profiles for case I and II CO ferroelectrics. In case I, the non-polar structure of Table~\ref{table4} coincides with the centrosymmetric structures of Table S1. Whereas in case II, the centrosymmetric structures of Table S2 correspond to the parent structure in Table~\ref{table4}.

In all 21 cases, the ferroelectric transition displays a coupling term of the form $Q_1Q^2_2$. In case I, non-polar to polar energy differences and volume changes are relatively small, $<100$~meV/atom and $<1.5$\%, respectively, since both non-polar and polar structures display local distortions around the metal sites and correspond to local minima (see Fig.~\ref{figure4}a). Except for 4 cases, the ferroelectric phase is the ground state (i.e. $\Delta E <0$). For case II, the parent to polar structural transitions imply large displacements of ions, and therefore, in several cases, there are significant energy gains ($>500$~meV/atom) and volume change ($>10$\%) between parent and polar phases. However, in this case, the switching process is not expected to occur through the parent structure, but rather due to charge rearrangement through the barrier (see Fig.~\ref{figure4}b). In case II, the polar structure is always lower than the parent structure.

The materials reported in Table~\ref{table4} correspond to new CO-induced ferroelectric candidates proposed in this work. Note that despite the fact that all polar structures are reported in the Materials Project, none of these materials has been previously proposed as CO-induced ferroelectrics. LiCr$_2$O$_4$ features a distorted spinel structure and has been predicted to be a half-metal~\cite{Lauer2004}. A related material, LiMn$_2$O$_4$, is a well-known cathode material for Li-ion batteries and displays CO~\cite{Rodriguez-Carvajal1998}. However, unfortunately, not much is known about the rest of the materials. We were not able to find experimental literature on these compounds, and we speculate that these materials have been predicted from first principles for Li battery applications. 

There are 7 materials with two metal ions. From these, 5 display CO on only one of the metal ions. In detail, Li$_4$TiCo$_5$O$_{12}$, Li$_5$Ti$_2$Fe$_3$O$_{10}$, Li$_2$MnV$_3$O$_8$, Li$_2$Mn$_3$VO$_8$ and MnAl(WO$_4$)$_2$ display CO on Co, Fe, Mn, V and W respectively. In all these cases, the secondary metal ion remains in the same valence state. In MnAl(WO$_4$)$_2$, the Mn valence changes from $3+$ in the non-polar structure to $2+$  in the polar phase. In addition, AlNi(WO$_4$)$_2$ and Li$_2$Fe(CoO$_3$)$_2$ display CO in both metal ions. The latter is originated by the fact that the local environments of the two different ions are connected; therefore, as one of the ions' octahedron volume increases, the other ion's octahedra decrease.

The Materials Project predicts 3 ferromagnetic cases, namely Li$_4$Co$_5$SnO$_{12}$, Li$_5$Ti$_2$Fe$_3$O$_{10}$, LiCr$_2$O$_4$, and 4 ferrimagnetic cases, namely Fe$_4$As$_5$O$_{13}$, Li$_7$Co$_5$O$_{12}$, Li(Fe$_2$O$_3$)$_4$, V$_3$(O$_2$F)$_2$, while the rest are predicted as antiferromagnetic. Therefore, all materials in Table~\ref{table4} are also multiferroic candidates. However, magnetic properties have been studied for a portion of the Materials Project database using collinear calculations~\cite{Horton2019}. Further calculations considering non-collinear alternative magnetic configurations are required to confirm the magnetic ground state structure. We leave these calculations for future work.

\section{Discussion} 

Table~\ref{table1} reports the results of the high throughput methodology. Our filtering criteria to screen for CO-induced ferroelectrics, namely (i) different Wyckoff positions in the non-polar (case I) or equal Wyckoff positions in the parent (case II)  structure, (ii) different valence states in the polar structure, and (iii) polar to nonpolar structural transition under the suppression of the spin degree of freedom, can be applied to any materials database. For instance, if the filtering criteria are applied to the ferroelectric database of Smidt et al.~\cite{Smidt2020} (including high and non-high quality $\sim400$ materials), the workflow leads to 10 magnetic materials with different valence states in the polar structure, 5 with different Wyckoff positions in the nonpolar structure (case I) and 5 with equal Wyckoff positions in the parent (case II). From these, 3 and 1 are already included in Tables S3 and S4, respectively, namely \textsl{Cm} Fe$_3$O$_4$, \textsl{Pmc2$_1$} Fe$_3$O$_4$, \textsl{P1} V$_7$O$_{13}$, and \textsl{R3m} Al(MoS$_2$)$_4$. Further relaxation of the remaining candidates, suppressing spin, leads to 2 additional CO-induced ferroelectric candidates, namely \textsl{C2/m} W$_3$O$_8$ in case I and \textsl{P2$_1$/c} Fe$_7$S$_8$ in case II. The existence of additional candidates in the ferroelectric database of Ref.~\cite{Smidt2020} arises due to methodological differences in the construction of the databases and the fact that the Materials Project is a dynamical database. For instance, Ref.~\cite{Smidt2020} relied on the BCS to obtain the symmetry pairs, whereas our work utilizes \texttt{Pyxtal}. In addition, the polarization calculations performed in Ref.~\cite{Smidt2020} require the entire adiabatic path to be insulating, including the parent structure. The latter significantly reduces the number of material candidates available to find CO ferroelectrics. 

\begin{table}[htp]
\caption{Number of resulting materials along the high throughput screening workflow of CO-induced ferroelectrics in the Materials Project database in terms of different categories determined by a given property or descriptor.}
\label{table1}
\begin{tabular}{lr}
\hline \hline
Criteria   & $\#$ Materials    \\  \hline
Materials Project & 150,000 \\
Polar &  29,830 \\
Polar insulating &  19,107 \\ 
Ferroelectric symmetry pairs & 945 \\
CO candidates (case I)&  80 \\ 
CO candidates (case II)& 67  \\ 

CO-induced FE (case I) &  9 \\ 
CO-induced FE (case II) &  12 \\ 
\hline
Total CO-induced FE &  21 \\ 

\hline \hline
\end{tabular}
\end{table}

The materials reported in Table~\ref{table4} correspond to our final list of predicted CO-induced ferroelectrics. Each of these materials corresponds to a representative of an entire family of CO-induced ferroelectrics with the same ferroelectric symmetry pair, i.e. possessing the same combination of structural symmetries and symmetry-breaking modes. Furthermore, the parent structures are hypothetical and are not necessarily accessible experimentally. Note that suppressing the spin in the ferromagnetic DFT calculation of the CO material will necessarily imply that metal ions at equivalent sites will share an electron. Therefore, the parent structures in Table~\ref{table4} are expected to display half-integer valence states and, therefore, be metallic. 

Our search strategy was able to recover previously known cases, such as LiFe$_2$F$_6$ and Fe$_3$O$_4$. Several other known CO ferroelectrics are reported in The Materials Project database, namely Ag$_2$BiO$_3$ (\textsl{Pnna, Pnn2}), BiMn$_2$O$_5$ (\textsl{Pbam}), La$_2$NiMnO$_6$ (\textsl{P2$_1$/c}), LuFe$_2$O$_4$ (\textsl{C2/m}). However, only Ag$_2$BiO$_3$ and LiFeF$_6$ are reported in both centrosymmetric and polar structures. Unfortunately, the valence states are not assigned to Ag$_2$BiO$_3$ \textsl{Pnn2}, and this case is missed by the workflow. The latter highlights the importance of extending the current implementation of the BVS method in \texttt{Pymatgen}. We leave this for future work.


Our results clearly distinguish between materials with and without a coupling term between the CO mode and the polar mode. Table~\ref{table4} reports the materials with a coupling term, whereas Tables S1 and S2 report all CO ferroelectric candidates with and without a coupling term. We note that in some cases, the polar distortion is dominated by ion displacement related to the CO, whereas in other cases, the polar distortion is dominated by other ion displacements. The electronic-lattice coupling is expected to vary between two extremes, with electrons being loosely and tightly bound to the lattice, and with faster and slower charge transfer between sites.

Our spin suppression method may also be used to screen for coupled ferroelectric ferromagnetic multiferroics. For all the materials reported with a ferromagnetic polar structure in the Materials Project, we can construct an artificial parent structure by artificially removing the spin degree of freedom in the DFT calculation. In all cases where the polar structure transforms to a non-polar structure, we can consider them as candidates. We left this for future work.

\section{Conclusions} 

In summary, we have performed a high-throughput screening of CO ferroelectrics in the Materials Project database. Among an initial list of 945 ferroelectric symmetry pairs described by a pair of non-polar and polar structures related by symmetry, we identify 147 CO ferroelectrics as those having different Wyckoff positions in the non-polar structure, equal Wyckoff positions in the parent structure, and different valence states in the polar structure. In addition, if the polar structure converges to a centrosymmetric structure when the spin is artificially suppressed from the DFT structural relaxation, the CO ferroelectric displays a coupling term between CO and polar distortions, and therefore corresponds to a CO-induced ferroelectric. Our workflow recovers 2 well-known CO-induced ferroelectrics and identifies 21 new CO-induced ferroelectric candidates in the Materials Project database. 

\bigskip

\section{\label{sec:level1} Acknowledgments}

We thank Karin M. Rabe and Se Young Park for valuable discussions. J.C. and S.E.R.-L. acknowledge support from ANID FONDECYT Regular grant number 1220986. J.C. and S.E.R.-L. also acknowledge support from the Abdus Salam International Center for Theoretical Physics (ICTP) through the Sandwich Training Educational Programme (2023-2025) and Associates Programme (2021-2026), respectively. Y.Q. was supported by the Office of Naval Research through N00014-21-1-210. Powered@NLHPC: This research was supported by the supercomputing infrastructure of the National Laboratory for High Performance Computing Chile (CCSS210001).


\begin{thebibliography}{79}%
\makeatletter
\providecommand \@ifxundefined [1]{%
 \@ifx{#1\undefined}
}%
\providecommand \@ifnum [1]{%
 \ifnum #1\expandafter \@firstoftwo
 \else \expandafter \@secondoftwo
 \fi
}%
\providecommand \@ifx [1]{%
 \ifx #1\expandafter \@firstoftwo
 \else \expandafter \@secondoftwo
 \fi
}%
\providecommand \natexlab [1]{#1}%
\providecommand \enquote  [1]{``#1''}%
\providecommand \bibnamefont  [1]{#1}%
\providecommand \bibfnamefont [1]{#1}%
\providecommand \citenamefont [1]{#1}%
\providecommand \href@noop [0]{\@secondoftwo}%
\providecommand \href [0]{\begingroup \@sanitize@url \@href}%
\providecommand \@href[1]{\@@startlink{#1}\@@href}%
\providecommand \@@href[1]{\endgroup#1\@@endlink}%
\providecommand \@sanitize@url [0]{\catcode `\\12\catcode `\$12\catcode
  `\&12\catcode `\#12\catcode `\^12\catcode `\_12\catcode `\%12\relax}%
\providecommand \@@startlink[1]{}%
\providecommand \@@endlink[0]{}%
\providecommand \url  [0]{\begingroup\@sanitize@url \@url }%
\providecommand \@url [1]{\endgroup\@href {#1}{\urlprefix }}%
\providecommand \urlprefix  [0]{URL }%
\providecommand \Eprint [0]{\href }%
\providecommand \doibase [0]{https://doi.org/}%
\providecommand \selectlanguage [0]{\@gobble}%
\providecommand \bibinfo  [0]{\@secondoftwo}%
\providecommand \bibfield  [0]{\@secondoftwo}%
\providecommand \translation [1]{[#1]}%
\providecommand \BibitemOpen [0]{}%
\providecommand \bibitemStop [0]{}%
\providecommand \bibitemNoStop [0]{.\EOS\space}%
\providecommand \EOS [0]{\spacefactor3000\relax}%
\providecommand \BibitemShut  [1]{\csname bibitem#1\endcsname}%
\let\auto@bib@innerbib\@empty
\bibitem [{\citenamefont {Rabe}\ \emph {et~al.}(2007)\citenamefont {Rabe},
  \citenamefont {Ahn},\ and\ \citenamefont {Triscone}}]{Rabe-book}%
  \BibitemOpen
  \bibfield  {author} {\bibinfo {author} {\bibfnamefont {K.~M.}\ \bibnamefont
  {Rabe}}, \bibinfo {author} {\bibfnamefont {C.~H.}\ \bibnamefont {Ahn}},\ and\
  \bibinfo {author} {\bibfnamefont {J.~M.}\ \bibnamefont {Triscone}},\
  }\href@noop {} {{\bibinfo {title} {{Physics of ferroelectrics: a modern
  perspective}}}}\ (\bibinfo  {publisher} {Springer},\ \bibinfo {year}
  {2007})\BibitemShut {NoStop}%
\bibitem [{\citenamefont {{Van Den Brink}}\ and\ \citenamefont
  {Khomskii}(2008)}]{VanDenBrink2008}%
  \BibitemOpen
  \bibfield  {author} {\bibinfo {author} {\bibfnamefont {J.}~\bibnamefont {{Van
  Den Brink}}}\ and\ \bibinfo {author} {\bibfnamefont {D.~I.}\ \bibnamefont
  {Khomskii}},\ }\bibfield  {title} {\bibinfo {title} {{Multiferroicity due to
  charge ordering}},\ }\href@noop {} {\bibfield  {journal} {\bibinfo  {journal}
  {Journal of Physics Condensed Matter}\ }\textbf {\bibinfo {volume} {20}}
  (\bibinfo {year} {2008})}\BibitemShut {NoStop}%
\bibitem [{\citenamefont {Picozzi}\ and\ \citenamefont
  {Ederer}(2009)}]{Picozzi2009}%
  \BibitemOpen
  \bibfield  {author} {\bibinfo {author} {\bibfnamefont {S.}~\bibnamefont
  {Picozzi}}\ and\ \bibinfo {author} {\bibfnamefont {C.}~\bibnamefont
  {Ederer}},\ }\bibfield  {title} {\bibinfo {title} {{First principles studies
  of multiferroic materials}},\ }\href
  {https://doi.org/10.1088/0953-8984/21/30/303201} {\bibfield  {journal}
  {\bibinfo  {journal} {Journal of Physics Condensed Matter}\ }\textbf
  {\bibinfo {volume} {21}},\ \bibinfo {pages} {303201} (\bibinfo {year}
  {2009})}\BibitemShut {NoStop}%
\bibitem [{\citenamefont {Khomskii}(2009)}]{Khomskii2009}%
  \BibitemOpen
  \bibfield  {author} {\bibinfo {author} {\bibfnamefont {D.}~\bibnamefont
  {Khomskii}},\ }\bibfield  {title} {\bibinfo {title} {{Classifying
  multiferroics: Mechanisms and effects}},\ }\href
  {https://doi.org/10.1103/physics.2.20} {\bibfield  {journal} {\bibinfo
  {journal} {Physics}\ }\textbf {\bibinfo {volume} {2}},\ \bibinfo {pages} {20}
  (\bibinfo {year} {2009})}\BibitemShut {NoStop}%
\bibitem [{\citenamefont {Spaldin}\ and\ \citenamefont
  {Fiebig}(2005)}]{Spaldin2005}%
  \BibitemOpen
  \bibfield  {author} {\bibinfo {author} {\bibfnamefont {N.~A.}\ \bibnamefont
  {Spaldin}}\ and\ \bibinfo {author} {\bibfnamefont {M.}~\bibnamefont
  {Fiebig}},\ }\bibfield  {title} {\bibinfo {title} {{The Renaissance of
  Magnetoelectric Multiferroics}},\ }\href@noop {} {\bibfield  {journal}
  {\bibinfo  {journal} {Science}\ }\textbf {\bibinfo {volume} {309}},\ \bibinfo
  {pages} {391} (\bibinfo {year} {2005})}\BibitemShut {NoStop}%
\bibitem [{\citenamefont {Fiebig}\ \emph {et~al.}(2016)\citenamefont {Fiebig},
  \citenamefont {Lottermoser}, \citenamefont {Meier},\ and\ \citenamefont
  {Trassin}}]{Fiebig2016}%
  \BibitemOpen
  \bibfield  {author} {\bibinfo {author} {\bibfnamefont {M.}~\bibnamefont
  {Fiebig}}, \bibinfo {author} {\bibfnamefont {T.}~\bibnamefont {Lottermoser}},
  \bibinfo {author} {\bibfnamefont {D.}~\bibnamefont {Meier}},\ and\ \bibinfo
  {author} {\bibfnamefont {M.}~\bibnamefont {Trassin}},\ }\bibfield  {title}
  {\bibinfo {title} {{The evolution of multiferroics}},\ }\bibfield  {journal}
  {\bibinfo  {journal} {Nature Reviews Materials}\ }\textbf {\bibinfo {volume}
  {1}},\ \href {https://doi.org/10.1038/natrevmats.2016.46}
  {10.1038/natrevmats.2016.46} (\bibinfo {year} {2016})\BibitemShut {NoStop}%
\bibitem [{\citenamefont {Spaldin}\ and\ \citenamefont
  {Ramesh}(2019)}]{Spaldin2019}%
  \BibitemOpen
  \bibfield  {author} {\bibinfo {author} {\bibfnamefont {N.~A.}\ \bibnamefont
  {Spaldin}}\ and\ \bibinfo {author} {\bibfnamefont {R.}~\bibnamefont
  {Ramesh}},\ }\bibfield  {title} {\bibinfo {title} {{Advances in
  magnetoelectric multiferroics}},\ }\href
  {https://doi.org/10.1038/s41563-018-0275-2} {\bibfield  {journal} {\bibinfo
  {journal} {Nature Materials}\ }\textbf {\bibinfo {volume} {18}},\ \bibinfo
  {pages} {203} (\bibinfo {year} {2019})}\BibitemShut {NoStop}%
\bibitem [{\citenamefont {Eduardo}\ \emph {et~al.}(2014)\citenamefont
  {Eduardo}, \citenamefont {Neto}, \citenamefont {Aynajian}, \citenamefont
  {Frano}, \citenamefont {Comin}, \citenamefont {Schierle}, \citenamefont
  {Weschke}, \citenamefont {Gyenis}, \citenamefont {Wen}, \citenamefont
  {Schneeloch}, \citenamefont {Xu}, \citenamefont {Ono}, \citenamefont {Gu},
  \citenamefont {Tacon},\ and\ \citenamefont {Yazdani}}]{Eduardo2014}%
  \BibitemOpen
  \bibfield  {author} {\bibinfo {author} {\bibfnamefont {H.}~\bibnamefont
  {Eduardo}}, \bibinfo {author} {\bibfnamefont {S.}~\bibnamefont {Neto}},
  \bibinfo {author} {\bibfnamefont {P.}~\bibnamefont {Aynajian}}, \bibinfo
  {author} {\bibfnamefont {A.}~\bibnamefont {Frano}}, \bibinfo {author}
  {\bibfnamefont {R.}~\bibnamefont {Comin}}, \bibinfo {author} {\bibfnamefont
  {E.}~\bibnamefont {Schierle}}, \bibinfo {author} {\bibfnamefont
  {E.}~\bibnamefont {Weschke}}, \bibinfo {author} {\bibfnamefont
  {A.}~\bibnamefont {Gyenis}}, \bibinfo {author} {\bibfnamefont
  {J.}~\bibnamefont {Wen}}, \bibinfo {author} {\bibfnamefont {J.}~\bibnamefont
  {Schneeloch}}, \bibinfo {author} {\bibfnamefont {Z.}~\bibnamefont {Xu}},
  \bibinfo {author} {\bibfnamefont {S.}~\bibnamefont {Ono}}, \bibinfo {author}
  {\bibfnamefont {G.}~\bibnamefont {Gu}}, \bibinfo {author} {\bibfnamefont
  {M.~L.}\ \bibnamefont {Tacon}},\ and\ \bibinfo {author} {\bibfnamefont
  {A.}~\bibnamefont {Yazdani}},\ }\bibfield  {title} {\bibinfo {title}
  {{Ubiquitous Interplay Between Charge Ordering and High-Temperature
  Superconductivity in Cuprates}},\ }\href@noop {} {\bibfield  {journal}
  {\bibinfo  {journal} {Science}\ }\textbf {\bibinfo {volume} {343}},\ \bibinfo
  {pages} {393} (\bibinfo {year} {2014})}\BibitemShut {NoStop}%
\bibitem [{\citenamefont {Zheng}\ \emph {et~al.}(2022)\citenamefont {Zheng},
  \citenamefont {Wu}, \citenamefont {Yang}, \citenamefont {Nie}, \citenamefont
  {Shan}, \citenamefont {Sun}, \citenamefont {Song}, \citenamefont {Yu},
  \citenamefont {Li}, \citenamefont {Zhao}, \citenamefont {Li}, \citenamefont
  {Kang}, \citenamefont {Zhou}, \citenamefont {Liu}, \citenamefont {Xiang},
  \citenamefont {Ying}, \citenamefont {Wang}, \citenamefont {Wu},\ and\
  \citenamefont {Chen}}]{Zheng2022}%
  \BibitemOpen
  \bibfield  {author} {\bibinfo {author} {\bibfnamefont {L.}~\bibnamefont
  {Zheng}}, \bibinfo {author} {\bibfnamefont {Z.}~\bibnamefont {Wu}}, \bibinfo
  {author} {\bibfnamefont {Y.}~\bibnamefont {Yang}}, \bibinfo {author}
  {\bibfnamefont {L.}~\bibnamefont {Nie}}, \bibinfo {author} {\bibfnamefont
  {M.}~\bibnamefont {Shan}}, \bibinfo {author} {\bibfnamefont {K.}~\bibnamefont
  {Sun}}, \bibinfo {author} {\bibfnamefont {D.}~\bibnamefont {Song}}, \bibinfo
  {author} {\bibfnamefont {F.}~\bibnamefont {Yu}}, \bibinfo {author}
  {\bibfnamefont {J.}~\bibnamefont {Li}}, \bibinfo {author} {\bibfnamefont
  {D.}~\bibnamefont {Zhao}}, \bibinfo {author} {\bibfnamefont {S.}~\bibnamefont
  {Li}}, \bibinfo {author} {\bibfnamefont {B.}~\bibnamefont {Kang}}, \bibinfo
  {author} {\bibfnamefont {Y.}~\bibnamefont {Zhou}}, \bibinfo {author}
  {\bibfnamefont {K.}~\bibnamefont {Liu}}, \bibinfo {author} {\bibfnamefont
  {Z.}~\bibnamefont {Xiang}}, \bibinfo {author} {\bibfnamefont
  {J.}~\bibnamefont {Ying}}, \bibinfo {author} {\bibfnamefont {Z.}~\bibnamefont
  {Wang}}, \bibinfo {author} {\bibfnamefont {T.}~\bibnamefont {Wu}},\ and\
  \bibinfo {author} {\bibfnamefont {X.}~\bibnamefont {Chen}},\ }\bibfield
  {title} {\bibinfo {title} {{Emergent charge order in pressurized kagome
  superconductor CsV3Sb5}},\ }\href
  {https://doi.org/10.1038/s41586-022-05351-3} {\bibfield  {journal} {\bibinfo
  {journal} {Nature}\ }\textbf {\bibinfo {volume} {611}},\ \bibinfo {pages}
  {682} (\bibinfo {year} {2022})}\BibitemShut {NoStop}%
\bibitem [{\citenamefont {Şen}\ \emph {et~al.}(2007)\citenamefont {Şen},
  \citenamefont {Alvarez},\ and\ \citenamefont {Dagotto}}]{Sen2007}%
  \BibitemOpen
  \bibfield  {author} {\bibinfo {author} {\bibfnamefont {C.}~\bibnamefont
  {Şen}}, \bibinfo {author} {\bibfnamefont {G.}~\bibnamefont {Alvarez}},\ and\
  \bibinfo {author} {\bibfnamefont {E.}~\bibnamefont {Dagotto}},\ }\bibfield
  {title} {\bibinfo {title} {{Competing Ferromagnetic and Charge-Ordered States
  in Models for Manganites: The Origin of the Colossal Magnetoresistance
  Effect}},\ }\href {https://doi.org/10.1103/PhysRevLett.98.127202} {\bibfield
  {journal} {\bibinfo  {journal} {Physical Review Letters}\ }\textbf {\bibinfo
  {volume} {98}},\ \bibinfo {pages} {1} (\bibinfo {year} {2007})}\BibitemShut
  {NoStop}%
\bibitem [{\citenamefont {Kawakami}\ \emph {et~al.}(2010)\citenamefont
  {Kawakami}, \citenamefont {Fukatsu}, \citenamefont {Sakurai}, \citenamefont
  {Unno}, \citenamefont {Itoh}, \citenamefont {Iwai}, \citenamefont {Sasaki},
  \citenamefont {Yamamoto}, \citenamefont {Yakushi},\ and\ \citenamefont
  {Yonemitsu}}]{Kawakami2010}%
  \BibitemOpen
  \bibfield  {author} {\bibinfo {author} {\bibfnamefont {Y.}~\bibnamefont
  {Kawakami}}, \bibinfo {author} {\bibfnamefont {T.}~\bibnamefont {Fukatsu}},
  \bibinfo {author} {\bibfnamefont {Y.}~\bibnamefont {Sakurai}}, \bibinfo
  {author} {\bibfnamefont {H.}~\bibnamefont {Unno}}, \bibinfo {author}
  {\bibfnamefont {H.}~\bibnamefont {Itoh}}, \bibinfo {author} {\bibfnamefont
  {S.}~\bibnamefont {Iwai}}, \bibinfo {author} {\bibfnamefont {T.}~\bibnamefont
  {Sasaki}}, \bibinfo {author} {\bibfnamefont {K.}~\bibnamefont {Yamamoto}},
  \bibinfo {author} {\bibfnamefont {K.}~\bibnamefont {Yakushi}},\ and\ \bibinfo
  {author} {\bibfnamefont {K.}~\bibnamefont {Yonemitsu}},\ }\bibfield  {title}
  {\bibinfo {title} {{Early-stage dynamics of light-matter interaction leading
  to the insulator-to-metal transition in a charge ordered organic crystal}},\
  }\href {https://doi.org/10.1103/PhysRevLett.105.246402} {\bibfield  {journal}
  {\bibinfo  {journal} {Physical Review Letters}\ }\textbf {\bibinfo {volume}
  {105}},\ \bibinfo {pages} {2} (\bibinfo {year} {2010})}\BibitemShut {NoStop}%
\bibitem [{\citenamefont {Alonso}\ \emph {et~al.}(2000)\citenamefont {Alonso},
  \citenamefont {Mart{\'{i}}nez-Lope}, \citenamefont {Casais},\ and\
  \citenamefont {Garc{\'{i}}a-Mu{\~{n}}oz}}]{Alonso2000}%
  \BibitemOpen
  \bibfield  {author} {\bibinfo {author} {\bibfnamefont {J.}~\bibnamefont
  {Alonso}}, \bibinfo {author} {\bibfnamefont {M.}~\bibnamefont
  {Mart{\'{i}}nez-Lope}}, \bibinfo {author} {\bibfnamefont {M.}~\bibnamefont
  {Casais}},\ and\ \bibinfo {author} {\bibfnamefont {J.}~\bibnamefont
  {Garc{\'{i}}a-Mu{\~{n}}oz}},\ }\bibfield  {title} {\bibinfo {title}
  {{Room-temperature monoclinic distortion due to charge disproportionation in
  perovskites with small rare-earth cations Y, Er, Tm, Yb, and Lu): A neutron
  diffraction study}},\ }\href {https://doi.org/10.1103/PhysRevB.61.1756}
  {\bibfield  {journal} {\bibinfo  {journal} {Physical Review B - Condensed
  Matter and Materials Physics}\ }\textbf {\bibinfo {volume} {61}},\ \bibinfo
  {pages} {1756} (\bibinfo {year} {2000})}\BibitemShut {NoStop}%
\bibitem [{\citenamefont {Ovsyannikov}\ \emph {et~al.}(2016)\citenamefont
  {Ovsyannikov}, \citenamefont {Bykov}, \citenamefont {Bykova}, \citenamefont
  {Kozlenko}, \citenamefont {Tsirlin}, \citenamefont {Karkin}, \citenamefont
  {Shchennikov}, \citenamefont {Kichanov}, \citenamefont {Gou}, \citenamefont
  {Abakumov}, \citenamefont {Egoavil}, \citenamefont {Verbeeck}, \citenamefont
  {McCammon}, \citenamefont {Dyadkin}, \citenamefont {Chernyshov},
  \citenamefont {{Van Smaalen}},\ and\ \citenamefont
  {Dubrovinsky}}]{Ovsyannikov2016}%
  \BibitemOpen
  \bibfield  {author} {\bibinfo {author} {\bibfnamefont {S.~V.}\ \bibnamefont
  {Ovsyannikov}}, \bibinfo {author} {\bibfnamefont {M.}~\bibnamefont {Bykov}},
  \bibinfo {author} {\bibfnamefont {E.}~\bibnamefont {Bykova}}, \bibinfo
  {author} {\bibfnamefont {D.~P.}\ \bibnamefont {Kozlenko}}, \bibinfo {author}
  {\bibfnamefont {A.~A.}\ \bibnamefont {Tsirlin}}, \bibinfo {author}
  {\bibfnamefont {A.~E.}\ \bibnamefont {Karkin}}, \bibinfo {author}
  {\bibfnamefont {V.~V.}\ \bibnamefont {Shchennikov}}, \bibinfo {author}
  {\bibfnamefont {S.~E.}\ \bibnamefont {Kichanov}}, \bibinfo {author}
  {\bibfnamefont {H.}~\bibnamefont {Gou}}, \bibinfo {author} {\bibfnamefont
  {A.~M.}\ \bibnamefont {Abakumov}}, \bibinfo {author} {\bibfnamefont
  {R.}~\bibnamefont {Egoavil}}, \bibinfo {author} {\bibfnamefont
  {J.}~\bibnamefont {Verbeeck}}, \bibinfo {author} {\bibfnamefont
  {C.}~\bibnamefont {McCammon}}, \bibinfo {author} {\bibfnamefont
  {V.}~\bibnamefont {Dyadkin}}, \bibinfo {author} {\bibfnamefont
  {D.}~\bibnamefont {Chernyshov}}, \bibinfo {author} {\bibfnamefont
  {S.}~\bibnamefont {{Van Smaalen}}},\ and\ \bibinfo {author} {\bibfnamefont
  {L.~S.}\ \bibnamefont {Dubrovinsky}},\ }\bibfield  {title} {\bibinfo {title}
  {{Charge-ordering transition in iron oxide Fe4O5 involving competing dimer
  and trimer formation}},\ }\href {https://doi.org/10.1038/nchem.2478}
  {\bibfield  {journal} {\bibinfo  {journal} {Nature Chemistry}\ }\textbf
  {\bibinfo {volume} {8}},\ \bibinfo {pages} {501} (\bibinfo {year}
  {2016})}\BibitemShut {NoStop}%
\bibitem [{\citenamefont {He}\ \emph {et~al.}(2018)\citenamefont {He},
  \citenamefont {{Di Sante}}, \citenamefont {Li}, \citenamefont {Chen},
  \citenamefont {Rondinelli},\ and\ \citenamefont {Franchini}}]{He2018}%
  \BibitemOpen
  \bibfield  {author} {\bibinfo {author} {\bibfnamefont {J.}~\bibnamefont
  {He}}, \bibinfo {author} {\bibfnamefont {D.}~\bibnamefont {{Di Sante}}},
  \bibinfo {author} {\bibfnamefont {R.}~\bibnamefont {Li}}, \bibinfo {author}
  {\bibfnamefont {X.~Q.}\ \bibnamefont {Chen}}, \bibinfo {author}
  {\bibfnamefont {J.~M.}\ \bibnamefont {Rondinelli}},\ and\ \bibinfo {author}
  {\bibfnamefont {C.}~\bibnamefont {Franchini}},\ }\bibfield  {title} {\bibinfo
  {title} {{Tunable metal-insulator transition, Rashba effect and Weyl Fermions
  in a relativistic charge-ordered ferroelectric oxide}},\ }\href
  {https://doi.org/10.1038/s41467-017-02814-4} {\bibfield  {journal} {\bibinfo
  {journal} {Nature Communications}\ }\textbf {\bibinfo {volume} {9}},\
  \bibinfo {pages} {492} (\bibinfo {year} {2018})}\BibitemShut {NoStop}%
\bibitem [{\citenamefont {Li}\ \emph {et~al.}(2011)\citenamefont {Li},
  \citenamefont {Yao}, \citenamefont {Gao}, \citenamefont {Sun},\ and\
  \citenamefont {Li}}]{Li2011}%
  \BibitemOpen
  \bibfield  {author} {\bibinfo {author} {\bibfnamefont {N.}~\bibnamefont
  {Li}}, \bibinfo {author} {\bibfnamefont {K.}~\bibnamefont {Yao}}, \bibinfo
  {author} {\bibfnamefont {G.}~\bibnamefont {Gao}}, \bibinfo {author}
  {\bibfnamefont {Z.}~\bibnamefont {Sun}},\ and\ \bibinfo {author}
  {\bibfnamefont {L.}~\bibnamefont {Li}},\ }\bibfield  {title} {\bibinfo
  {title} {{Charge, orbital and spin ordering in multiferroic BiMn2O 5: Density
  functional theory calculations}},\ }\href
  {https://doi.org/10.1039/c0cp02252g} {\bibfield  {journal} {\bibinfo
  {journal} {Physical Chemistry Chemical Physics}\ }\textbf {\bibinfo {volume}
  {13}},\ \bibinfo {pages} {9418} (\bibinfo {year} {2011})}\BibitemShut
  {NoStop}%
\bibitem [{\citenamefont {Balachandran}\ and\ \citenamefont
  {Rondinelli}(2013)}]{Balachandran2013}%
  \BibitemOpen
  \bibfield  {author} {\bibinfo {author} {\bibfnamefont {P.~V.}\ \bibnamefont
  {Balachandran}}\ and\ \bibinfo {author} {\bibfnamefont {J.~M.}\ \bibnamefont
  {Rondinelli}},\ }\bibfield  {title} {\bibinfo {title} {{Interplay of
  octahedral rotations and breathing distortions in charge-ordering perovskite
  oxides}},\ }\href {https://doi.org/10.1103/PhysRevB.88.054101} {\bibfield
  {journal} {\bibinfo  {journal} {Physical Review B - Condensed Matter and
  Materials Physics}\ }\textbf {\bibinfo {volume} {88}},\ \bibinfo {pages} {1}
  (\bibinfo {year} {2013})}\BibitemShut {NoStop}%
\bibitem [{\citenamefont {Jain}\ \emph {et~al.}(2013)\citenamefont {Jain},
  \citenamefont {Ong}, \citenamefont {Hautier}, \citenamefont {Chen},
  \citenamefont {Richards}, \citenamefont {Dacek}, \citenamefont {Cholia},
  \citenamefont {Gunter}, \citenamefont {Skinner}, \citenamefont {Ceder},\ and\
  \citenamefont {Persson}}]{Jain2013}%
  \BibitemOpen
  \bibfield  {author} {\bibinfo {author} {\bibfnamefont {A.}~\bibnamefont
  {Jain}}, \bibinfo {author} {\bibfnamefont {S.~P.}\ \bibnamefont {Ong}},
  \bibinfo {author} {\bibfnamefont {G.}~\bibnamefont {Hautier}}, \bibinfo
  {author} {\bibfnamefont {W.}~\bibnamefont {Chen}}, \bibinfo {author}
  {\bibfnamefont {W.~D.}\ \bibnamefont {Richards}}, \bibinfo {author}
  {\bibfnamefont {S.}~\bibnamefont {Dacek}}, \bibinfo {author} {\bibfnamefont
  {S.}~\bibnamefont {Cholia}}, \bibinfo {author} {\bibfnamefont
  {D.}~\bibnamefont {Gunter}}, \bibinfo {author} {\bibfnamefont
  {D.}~\bibnamefont {Skinner}}, \bibinfo {author} {\bibfnamefont
  {G.}~\bibnamefont {Ceder}},\ and\ \bibinfo {author} {\bibfnamefont {K.~A.}\
  \bibnamefont {Persson}},\ }\bibfield  {title} {\bibinfo {title} {{Commentary:
  The materials project: A materials genome approach to accelerating materials
  innovation}},\ }\href@noop {} {\bibfield  {journal} {\bibinfo  {journal} {APL
  Materials}\ }\textbf {\bibinfo {volume} {1}} (\bibinfo {year}
  {2013})}\BibitemShut {NoStop}%
\bibitem [{\citenamefont {Kresse}\ and\ \citenamefont
  {Furthm{\"{u}}ller}(1996)}]{Kresse1996}%
  \BibitemOpen
  \bibfield  {author} {\bibinfo {author} {\bibfnamefont {G.}~\bibnamefont
  {Kresse}}\ and\ \bibinfo {author} {\bibfnamefont {J.}~\bibnamefont
  {Furthm{\"{u}}ller}},\ }\bibfield  {title} {\bibinfo {title} {{Efficiency of
  ab-initio total energy calculations for metals and semiconductors using a
  plane-wave basis set}},\ }\href
  {https://doi.org/10.1016/0927-0256(96)00008-0} {\bibfield  {journal}
  {\bibinfo  {journal} {Computational Materials Science}\ }\textbf {\bibinfo
  {volume} {6}},\ \bibinfo {pages} {15} (\bibinfo {year} {1996})}\BibitemShut
  {NoStop}%
\bibitem [{\citenamefont {Perdew}\ \emph {et~al.}(1996)\citenamefont {Perdew},
  \citenamefont {Burke},\ and\ \citenamefont {Ernzerhof}}]{Perdew1996}%
  \BibitemOpen
  \bibfield  {author} {\bibinfo {author} {\bibfnamefont {J.~P.}\ \bibnamefont
  {Perdew}}, \bibinfo {author} {\bibfnamefont {K.}~\bibnamefont {Burke}},\ and\
  \bibinfo {author} {\bibfnamefont {M.}~\bibnamefont {Ernzerhof}},\ }\bibfield
  {title} {\bibinfo {title} {{Generalized Gradient Approximation Made
  Simple}},\ }\href@noop {} {\bibfield  {journal} {\bibinfo  {journal}
  {Physical Review Letters}\ }\textbf {\bibinfo {volume} {18}},\ \bibinfo
  {pages} {3865} (\bibinfo {year} {1996})}\BibitemShut {NoStop}%
\bibitem [{\citenamefont {Dudarev}\ and\ \citenamefont
  {Botton}(1998)}]{Dudarev1998}%
  \BibitemOpen
  \bibfield  {author} {\bibinfo {author} {\bibfnamefont {S.}~\bibnamefont
  {Dudarev}}\ and\ \bibinfo {author} {\bibfnamefont {G.}~\bibnamefont
  {Botton}},\ }\bibfield  {title} {\bibinfo {title} {{Electron-energy-loss
  spectra and the structural stability of nickel oxide: An LSDA+U study}},\
  }\href {https://doi.org/10.1103/PhysRevB.57.1505} {\bibfield  {journal}
  {\bibinfo  {journal} {Physical Review B - Condensed Matter and Materials
  Physics}\ }\textbf {\bibinfo {volume} {57}},\ \bibinfo {pages} {1505}
  (\bibinfo {year} {1998})}\BibitemShut {NoStop}%
\bibitem [{\citenamefont {Resta}(1994)}]{Resta1994}%
  \BibitemOpen
  \bibfield  {author} {\bibinfo {author} {\bibfnamefont {R.}~\bibnamefont
  {Resta}},\ }\bibfield  {title} {\bibinfo {title} {{Macroscopic polarization
  in crystalline dielectrics: the geoirIetric phase approach}},\ }\href@noop {}
  {\bibfield  {journal} {\bibinfo  {journal} {Reviews of Modern Physics}\
  }\textbf {\bibinfo {volume} {66}},\ \bibinfo {pages} {899} (\bibinfo {year}
  {1994})}\BibitemShut {NoStop}%
\bibitem [{\citenamefont {Wang}\ \emph {et~al.}(2006)\citenamefont {Wang},
  \citenamefont {Maxisch},\ and\ \citenamefont {Ceder}}]{Wang2006a}%
  \BibitemOpen
  \bibfield  {author} {\bibinfo {author} {\bibfnamefont {L.}~\bibnamefont
  {Wang}}, \bibinfo {author} {\bibfnamefont {T.}~\bibnamefont {Maxisch}},\ and\
  \bibinfo {author} {\bibfnamefont {G.}~\bibnamefont {Ceder}},\ }\bibfield
  {title} {\bibinfo {title} {{Oxidation energies of transition metal oxides
  within the GGA+U framework}},\ }\href@noop {} {\bibfield  {journal} {\bibinfo
   {journal} {Physical Review B - Condensed Matter and Materials Physics}\
  }\textbf {\bibinfo {volume} {73}},\ \bibinfo {pages} {195107} (\bibinfo
  {year} {2006})}\BibitemShut {NoStop}%
\bibitem [{\citenamefont {Jain}\ \emph
  {et~al.}(2011{\natexlab{a}})\citenamefont {Jain}, \citenamefont {Hautier},
  \citenamefont {Ong}, \citenamefont {Moore}, \citenamefont {Fischer},
  \citenamefont {Persson},\ and\ \citenamefont {Ceder}}]{Jain2011a}%
  \BibitemOpen
  \bibfield  {author} {\bibinfo {author} {\bibfnamefont {A.}~\bibnamefont
  {Jain}}, \bibinfo {author} {\bibfnamefont {G.}~\bibnamefont {Hautier}},
  \bibinfo {author} {\bibfnamefont {S.~P.}\ \bibnamefont {Ong}}, \bibinfo
  {author} {\bibfnamefont {C.~J.}\ \bibnamefont {Moore}}, \bibinfo {author}
  {\bibfnamefont {C.~C.}\ \bibnamefont {Fischer}}, \bibinfo {author}
  {\bibfnamefont {K.~A.}\ \bibnamefont {Persson}},\ and\ \bibinfo {author}
  {\bibfnamefont {G.}~\bibnamefont {Ceder}},\ }\bibfield  {title} {\bibinfo
  {title} {{Formation enthalpies by mixing GGA and GGA + U calculations}},\
  }\href@noop {} {\bibfield  {journal} {\bibinfo  {journal} {Physical Review B
  - Condensed Matter and Materials Physics}\ }\textbf {\bibinfo {volume}
  {84}},\ \bibinfo {pages} {045115} (\bibinfo {year}
  {2011}{\natexlab{a}})}\BibitemShut {NoStop}%
\bibitem [{\citenamefont {Zagorac}\ \emph {et~al.}(2019)\citenamefont
  {Zagorac}, \citenamefont {Muller}, \citenamefont {Ruehl}, \citenamefont
  {Zagorac},\ and\ \citenamefont {Rehme}}]{Zagorac2019}%
  \BibitemOpen
  \bibfield  {author} {\bibinfo {author} {\bibfnamefont {D.}~\bibnamefont
  {Zagorac}}, \bibinfo {author} {\bibfnamefont {H.}~\bibnamefont {Muller}},
  \bibinfo {author} {\bibfnamefont {S.}~\bibnamefont {Ruehl}}, \bibinfo
  {author} {\bibfnamefont {J.}~\bibnamefont {Zagorac}},\ and\ \bibinfo {author}
  {\bibfnamefont {S.}~\bibnamefont {Rehme}},\ }\bibfield  {title} {\bibinfo
  {title} {{Recent developments in the Inorganic Crystal Structure Database:
  Theoretical crystal structure data and related features}},\ }\href
  {https://doi.org/10.1107/S160057671900997X} {\bibfield  {journal} {\bibinfo
  {journal} {Journal of Applied Crystallography}\ }\textbf {\bibinfo {volume}
  {52}},\ \bibinfo {pages} {918} (\bibinfo {year} {2019})}\BibitemShut
  {NoStop}%
\bibitem [{\citenamefont {Ye}\ \emph {et~al.}(2018)\citenamefont {Ye},
  \citenamefont {Chen}, \citenamefont {Dwaraknath}, \citenamefont {Jain},
  \citenamefont {Ong},\ and\ \citenamefont {Persson}}]{Ye2018}%
  \BibitemOpen
  \bibfield  {author} {\bibinfo {author} {\bibfnamefont {W.}~\bibnamefont
  {Ye}}, \bibinfo {author} {\bibfnamefont {C.}~\bibnamefont {Chen}}, \bibinfo
  {author} {\bibfnamefont {S.}~\bibnamefont {Dwaraknath}}, \bibinfo {author}
  {\bibfnamefont {A.}~\bibnamefont {Jain}}, \bibinfo {author} {\bibfnamefont
  {S.~P.}\ \bibnamefont {Ong}},\ and\ \bibinfo {author} {\bibfnamefont {K.~A.}\
  \bibnamefont {Persson}},\ }\bibfield  {title} {\bibinfo {title} {{Harnessing
  the Materials Project for machine-learning and accelerated discovery}},\
  }\href {https://doi.org/10.1557/mrs.2018.202} {\bibfield  {journal} {\bibinfo
   {journal} {MRS Bulletin}\ }\textbf {\bibinfo {volume} {43}},\ \bibinfo
  {pages} {664} (\bibinfo {year} {2018})}\BibitemShut {NoStop}%
\bibitem [{\citenamefont {Fredericks}\ \emph {et~al.}(2021)\citenamefont
  {Fredericks}, \citenamefont {Parrish}, \citenamefont {Sayre},\ and\
  \citenamefont {Zhu}}]{Fredericks2021}%
  \BibitemOpen
  \bibfield  {author} {\bibinfo {author} {\bibfnamefont {S.}~\bibnamefont
  {Fredericks}}, \bibinfo {author} {\bibfnamefont {K.}~\bibnamefont {Parrish}},
  \bibinfo {author} {\bibfnamefont {D.}~\bibnamefont {Sayre}},\ and\ \bibinfo
  {author} {\bibfnamefont {Q.}~\bibnamefont {Zhu}},\ }\bibfield  {title}
  {\bibinfo {title} {{PyXtal: A Python library for crystal structure generation
  and symmetry analysis}},\ }\href@noop {} {\bibfield  {journal} {\bibinfo
  {journal} {Computer Physics Communications}\ }\textbf {\bibinfo {volume}
  {261}},\ \bibinfo {pages} {107810} (\bibinfo {year} {2021})}\BibitemShut
  {NoStop}%
\bibitem [{\citenamefont {Jain}\ \emph
  {et~al.}(2011{\natexlab{b}})\citenamefont {Jain}, \citenamefont {Hautier},
  \citenamefont {Moore}, \citenamefont {{Ping Ong}}, \citenamefont {Fischer},
  \citenamefont {Mueller}, \citenamefont {Persson},\ and\ \citenamefont
  {Ceder}}]{Jain2011b}%
  \BibitemOpen
  \bibfield  {author} {\bibinfo {author} {\bibfnamefont {A.}~\bibnamefont
  {Jain}}, \bibinfo {author} {\bibfnamefont {G.}~\bibnamefont {Hautier}},
  \bibinfo {author} {\bibfnamefont {C.~J.}\ \bibnamefont {Moore}}, \bibinfo
  {author} {\bibfnamefont {S.}~\bibnamefont {{Ping Ong}}}, \bibinfo {author}
  {\bibfnamefont {C.~C.}\ \bibnamefont {Fischer}}, \bibinfo {author}
  {\bibfnamefont {T.}~\bibnamefont {Mueller}}, \bibinfo {author} {\bibfnamefont
  {K.~A.}\ \bibnamefont {Persson}},\ and\ \bibinfo {author} {\bibfnamefont
  {G.}~\bibnamefont {Ceder}},\ }\bibfield  {title} {\bibinfo {title} {{A
  high-throughput infrastructure for density functional theory calculations}},\
  }\href {https://doi.org/10.1016/j.commatsci.2011.02.023} {\bibfield
  {journal} {\bibinfo  {journal} {Computational Materials Science}\ }\textbf
  {\bibinfo {volume} {50}},\ \bibinfo {pages} {2295} (\bibinfo {year}
  {2011}{\natexlab{b}})}\BibitemShut {NoStop}%
\bibitem [{\citenamefont {O'Keeffe}\ and\ \citenamefont
  {Brese}(1991)}]{OKeeffe1991}%
  \BibitemOpen
  \bibfield  {author} {\bibinfo {author} {\bibfnamefont {M.}~\bibnamefont
  {O'Keeffe}}\ and\ \bibinfo {author} {\bibfnamefont {N.~E.}\ \bibnamefont
  {Brese}},\ }\bibfield  {title} {\bibinfo {title} {{Atom Sizes and Bond
  Lengths in Molecules and Crystals}},\ }\href
  {https://doi.org/10.1021/ja00009a002} {\bibfield  {journal} {\bibinfo
  {journal} {Journal of the American Chemical Society}\ }\textbf {\bibinfo
  {volume} {113}},\ \bibinfo {pages} {3226} (\bibinfo {year}
  {1991})}\BibitemShut {NoStop}%
\bibitem [{\citenamefont {Capillas}\ \emph {et~al.}(2011)\citenamefont
  {Capillas}, \citenamefont {Tasci}, \citenamefont {{De La Flor}},
  \citenamefont {Orobengoa}, \citenamefont {Perez-Mato},\ and\ \citenamefont
  {Aroyo}}]{Capillas2011a}%
  \BibitemOpen
  \bibfield  {author} {\bibinfo {author} {\bibfnamefont {C.}~\bibnamefont
  {Capillas}}, \bibinfo {author} {\bibfnamefont {E.~S.}\ \bibnamefont {Tasci}},
  \bibinfo {author} {\bibfnamefont {G.}~\bibnamefont {{De La Flor}}}, \bibinfo
  {author} {\bibfnamefont {D.}~\bibnamefont {Orobengoa}}, \bibinfo {author}
  {\bibfnamefont {J.~M.}\ \bibnamefont {Perez-Mato}},\ and\ \bibinfo {author}
  {\bibfnamefont {M.~I.}\ \bibnamefont {Aroyo}},\ }\bibfield  {title} {\bibinfo
  {title} {{A new computer tool at the Bilbao Crystallographic Server to detect
  and characterize pseudosymmetry}},\ }\href
  {https://doi.org/10.1524/zkri.2011.1321} {\bibfield  {journal} {\bibinfo
  {journal} {Zeitschrift fur Kristallographie}\ }\textbf {\bibinfo {volume}
  {226}},\ \bibinfo {pages} {186} (\bibinfo {year} {2011})}\BibitemShut
  {NoStop}%
\bibitem [{\citenamefont {Stokes}\ \emph {et~al.}()\citenamefont {Stokes},
  \citenamefont {Hatch},\ and\ \citenamefont {Campbell}}]{isotropy}%
  \BibitemOpen
  \bibfield  {author} {\bibinfo {author} {\bibfnamefont {H.~T.}\ \bibnamefont
  {Stokes}}, \bibinfo {author} {\bibfnamefont {D.~M.}\ \bibnamefont {Hatch}},\
  and\ \bibinfo {author} {\bibfnamefont {B.~J.}\ \bibnamefont {Campbell}},\
  }\href@noop {} {\bibinfo {title} {{ISODISTORT, ISOTROPY Software Suite,
  iso.byu.edu}}}\BibitemShut {NoStop}%
\bibitem [{\citenamefont {Campbell}\ \emph {et~al.}(2006)\citenamefont
  {Campbell}, \citenamefont {Stokes}, \citenamefont {Tanner}, \citenamefont
  {Hatch}, \citenamefont {Campbell}, \citenamefont {Stokes}, \citenamefont
  {Tanner},\ and\ \citenamefont {Hatch}}]{Campbell2006}%
  \BibitemOpen
  \bibfield  {author} {\bibinfo {author} {\bibfnamefont {B.~J.}\ \bibnamefont
  {Campbell}}, \bibinfo {author} {\bibfnamefont {H.~T.}\ \bibnamefont
  {Stokes}}, \bibinfo {author} {\bibfnamefont {D.~E.}\ \bibnamefont {Tanner}},
  \bibinfo {author} {\bibfnamefont {D.~M.}\ \bibnamefont {Hatch}}, \bibinfo
  {author} {\bibfnamefont {B.~J.}\ \bibnamefont {Campbell}}, \bibinfo {author}
  {\bibfnamefont {H.~T.}\ \bibnamefont {Stokes}}, \bibinfo {author}
  {\bibfnamefont {D.~E.}\ \bibnamefont {Tanner}},\ and\ \bibinfo {author}
  {\bibfnamefont {D.~M.}\ \bibnamefont {Hatch}},\ }\bibfield  {title} {\bibinfo
  {title} {{ISODISPLACE : a web-based tool for exploring structural distortions
  ISODISPLACE : a web-based tool for exploring structural distortions}},\
  }\href {https://doi.org/10.1107/S0021889806014075} {\bibfield  {journal}
  {\bibinfo  {journal} {Applied Crystallography}\ }\textbf {\bibinfo {volume}
  {39}},\ \bibinfo {pages} {607} (\bibinfo {year} {2006})}\BibitemShut
  {NoStop}%
\bibitem [{\citenamefont {Hatch}\ and\ \citenamefont
  {Stokes}(2003)}]{Hatch2003}%
  \BibitemOpen
  \bibfield  {author} {\bibinfo {author} {\bibfnamefont {D.~M.}\ \bibnamefont
  {Hatch}}\ and\ \bibinfo {author} {\bibfnamefont {H.~T.}\ \bibnamefont
  {Stokes}},\ }\bibfield  {title} {\bibinfo {title} {{ INVARIANTS : program for
  obtaining a list of invariant polynomials of the order-parameter components
  associated with irreducible representations of a space group }},\ }\href
  {https://doi.org/10.1107/s0021889803005946} {\bibfield  {journal} {\bibinfo
  {journal} {Journal of Applied Crystallography}\ }\textbf {\bibinfo {volume}
  {36}},\ \bibinfo {pages} {951} (\bibinfo {year} {2003})}\BibitemShut
  {NoStop}%
\bibitem [{\citenamefont {Giovannetti}\ \emph
  {et~al.}(2009{\natexlab{a}})\citenamefont {Giovannetti}, \citenamefont
  {Kumar}, \citenamefont {Khomskii}, \citenamefont {Picozzi},\ and\
  \citenamefont {{Van Den Brink}}}]{Giovannetti2009a}%
  \BibitemOpen
  \bibfield  {author} {\bibinfo {author} {\bibfnamefont {G.}~\bibnamefont
  {Giovannetti}}, \bibinfo {author} {\bibfnamefont {S.}~\bibnamefont {Kumar}},
  \bibinfo {author} {\bibfnamefont {D.}~\bibnamefont {Khomskii}}, \bibinfo
  {author} {\bibfnamefont {S.}~\bibnamefont {Picozzi}},\ and\ \bibinfo {author}
  {\bibfnamefont {J.}~\bibnamefont {{Van Den Brink}}},\ }\bibfield  {title}
  {\bibinfo {title} {{Multiferroicity in rare-earth nickelates RNiO3}},\ }\href
  {https://doi.org/10.1103/PhysRevLett.103.156401} {\bibfield  {journal}
  {\bibinfo  {journal} {Physical Review Letters}\ }\textbf {\bibinfo {volume}
  {103}},\ \bibinfo {pages} {156401} (\bibinfo {year}
  {2009}{\natexlab{a}})}\BibitemShut {NoStop}%
\bibitem [{\citenamefont {Xin}\ \emph {et~al.}(2014)\citenamefont {Xin},
  \citenamefont {Wang}, \citenamefont {Sui}, \citenamefont {Wang},
  \citenamefont {Wang}, \citenamefont {Su}, \citenamefont {Zhao},\ and\
  \citenamefont {Liu}}]{Xin2014}%
  \BibitemOpen
  \bibfield  {author} {\bibinfo {author} {\bibfnamefont {C.}~\bibnamefont
  {Xin}}, \bibinfo {author} {\bibfnamefont {Y.}~\bibnamefont {Wang}}, \bibinfo
  {author} {\bibfnamefont {Y.}~\bibnamefont {Sui}}, \bibinfo {author}
  {\bibfnamefont {Y.}~\bibnamefont {Wang}}, \bibinfo {author} {\bibfnamefont
  {X.}~\bibnamefont {Wang}}, \bibinfo {author} {\bibfnamefont {Y.}~\bibnamefont
  {Su}}, \bibinfo {author} {\bibfnamefont {K.}~\bibnamefont {Zhao}},\ and\
  \bibinfo {author} {\bibfnamefont {X.}~\bibnamefont {Liu}},\ }\bibfield
  {title} {\bibinfo {title} {{Electronic, magnetic and ferroelectric properties
  of multiferroic TlNiO3: A first principles study}},\ }\href
  {https://doi.org/10.1016/j.commatsci.2013.09.060} {\bibfield  {journal}
  {\bibinfo  {journal} {Computational Materials Science}\ }\textbf {\bibinfo
  {volume} {82}},\ \bibinfo {pages} {191} (\bibinfo {year} {2014})}\BibitemShut
  {NoStop}%
\bibitem [{\citenamefont {Kim}\ \emph {et~al.}(2001)\citenamefont {Kim},
  \citenamefont {Demazeau}, \citenamefont {Alonso},\ and\ \citenamefont
  {Choy}}]{Kim2001}%
  \BibitemOpen
  \bibfield  {author} {\bibinfo {author} {\bibfnamefont {S.~J.}\ \bibnamefont
  {Kim}}, \bibinfo {author} {\bibfnamefont {G.}~\bibnamefont {Demazeau}},
  \bibinfo {author} {\bibfnamefont {J.~A.}\ \bibnamefont {Alonso}},\ and\
  \bibinfo {author} {\bibfnamefont {J.~H.}\ \bibnamefont {Choy}},\ }\bibfield
  {title} {\bibinfo {title} {{High pressure synthesis and crystal structure of
  a new Ni(III) perovskite: TlNiO3}},\ }\href
  {https://doi.org/10.1039/b007043m} {\bibfield  {journal} {\bibinfo  {journal}
  {Journal of Materials Chemistry}\ }\textbf {\bibinfo {volume} {11}},\
  \bibinfo {pages} {487} (\bibinfo {year} {2001})}\BibitemShut {NoStop}%
\bibitem [{\citenamefont {Lin}\ \emph {et~al.}(2017)\citenamefont {Lin},
  \citenamefont {Xu}, \citenamefont {Zhang}, \citenamefont {Zhang},
  \citenamefont {Liang},\ and\ \citenamefont {Dong}}]{Lin2017}%
  \BibitemOpen
  \bibfield  {author} {\bibinfo {author} {\bibfnamefont {L.~F.}\ \bibnamefont
  {Lin}}, \bibinfo {author} {\bibfnamefont {Q.~R.}\ \bibnamefont {Xu}},
  \bibinfo {author} {\bibfnamefont {Y.}~\bibnamefont {Zhang}}, \bibinfo
  {author} {\bibfnamefont {J.~J.}\ \bibnamefont {Zhang}}, \bibinfo {author}
  {\bibfnamefont {Y.~P.}\ \bibnamefont {Liang}},\ and\ \bibinfo {author}
  {\bibfnamefont {S.}~\bibnamefont {Dong}},\ }\bibfield  {title} {\bibinfo
  {title} {{Ferroelectric ferrimagnetic LiFe2 F6: Charge-ordering-mediated
  magnetoelectricity}},\ }\href
  {https://doi.org/10.1103/PhysRevMaterials.1.071401} {\bibfield  {journal}
  {\bibinfo  {journal} {Physical Review Materials}\ }\textbf {\bibinfo {volume}
  {1}},\ \bibinfo {pages} {2} (\bibinfo {year} {2017})}\BibitemShut {NoStop}%
\bibitem [{\citenamefont {Fourquet}\ \emph {et~al.}(1988)\citenamefont
  {Fourquet}, \citenamefont {{Le Samedi}},\ and\ \citenamefont
  {Calage}}]{Fourquet1988}%
  \BibitemOpen
  \bibfield  {author} {\bibinfo {author} {\bibfnamefont {J.~L.}\ \bibnamefont
  {Fourquet}}, \bibinfo {author} {\bibfnamefont {E.}~\bibnamefont {{Le
  Samedi}}},\ and\ \bibinfo {author} {\bibfnamefont {Y.}~\bibnamefont
  {Calage}},\ }\bibfield  {title} {\bibinfo {title} {{Le trirutile
  ordonn{\'{e}} LiFe2F6: Croissance cristalline et {\'{e}}tude structurale}},\
  }\href {https://doi.org/10.1016/0022-4596(88)90093-X} {\bibfield  {journal}
  {\bibinfo  {journal} {Journal of Solid State Chemistry}\ }\textbf {\bibinfo
  {volume} {77}},\ \bibinfo {pages} {84} (\bibinfo {year} {1988})}\BibitemShut
  {NoStop}%
\bibitem [{\citenamefont {Greenwood}\ \emph {et~al.}(1971)\citenamefont
  {Greenwood}, \citenamefont {Howe},\ and\ \citenamefont
  {Menil}}]{Greenwood1971}%
  \BibitemOpen
  \bibfield  {author} {\bibinfo {author} {\bibfnamefont {N.~N.}\ \bibnamefont
  {Greenwood}}, \bibinfo {author} {\bibfnamefont {A.~T.}\ \bibnamefont
  {Howe}},\ and\ \bibinfo {author} {\bibfnamefont {F.}~\bibnamefont {Menil}},\
  }\bibfield  {title} {\bibinfo {title} {{Mossbauer Studies of Order and
  Disorder in Rutile and Trirutile Com- pounds derived from FeF2}},\
  }\href@noop {} {\bibfield  {journal} {\bibinfo  {journal} {Inorg. Phys.
  Theor.}\ ,\ \bibinfo {pages} {2218}} (\bibinfo {year} {1971})}\BibitemShut
  {NoStop}%
\bibitem [{\citenamefont {Oberndorfer}\ \emph {et~al.}(2006)\citenamefont
  {Oberndorfer}, \citenamefont {Dinnebier}, \citenamefont {Ibberson},\ and\
  \citenamefont {Jansen}}]{Oberndorfer2006}%
  \BibitemOpen
  \bibfield  {author} {\bibinfo {author} {\bibfnamefont {C.~P.}\ \bibnamefont
  {Oberndorfer}}, \bibinfo {author} {\bibfnamefont {R.~E.}\ \bibnamefont
  {Dinnebier}}, \bibinfo {author} {\bibfnamefont {R.~M.}\ \bibnamefont
  {Ibberson}},\ and\ \bibinfo {author} {\bibfnamefont {M.}~\bibnamefont
  {Jansen}},\ }\bibfield  {title} {\bibinfo {title} {{Charge ordering in
  Ag2BiO3}},\ }\href {https://doi.org/10.1016/j.solidstatesciences.2006.02.002}
  {\bibfield  {journal} {\bibinfo  {journal} {Solid State Sciences}\ }\textbf
  {\bibinfo {volume} {8}},\ \bibinfo {pages} {267} (\bibinfo {year}
  {2006})}\BibitemShut {NoStop}%
\bibitem [{\citenamefont {Alexe}\ \emph {et~al.}(2009)\citenamefont {Alexe},
  \citenamefont {Ziese}, \citenamefont {Hesse}, \citenamefont {Esquinazi},
  \citenamefont {Yamauchi}, \citenamefont {Fukushima}, \citenamefont
  {Picozzi},\ and\ \citenamefont {G{\"{o}}sele}}]{Alexe2009}%
  \BibitemOpen
  \bibfield  {author} {\bibinfo {author} {\bibfnamefont {M.}~\bibnamefont
  {Alexe}}, \bibinfo {author} {\bibfnamefont {M.}~\bibnamefont {Ziese}},
  \bibinfo {author} {\bibfnamefont {D.}~\bibnamefont {Hesse}}, \bibinfo
  {author} {\bibfnamefont {P.}~\bibnamefont {Esquinazi}}, \bibinfo {author}
  {\bibfnamefont {K.}~\bibnamefont {Yamauchi}}, \bibinfo {author}
  {\bibfnamefont {T.}~\bibnamefont {Fukushima}}, \bibinfo {author}
  {\bibfnamefont {S.}~\bibnamefont {Picozzi}},\ and\ \bibinfo {author}
  {\bibfnamefont {U.}~\bibnamefont {G{\"{o}}sele}},\ }\bibfield  {title}
  {\bibinfo {title} {{Ferroelectric switching in multiferroic magnetite (Fe3O
  4) thin films}},\ }\href {https://doi.org/10.1002/adma.200901381} {\bibfield
  {journal} {\bibinfo  {journal} {Advanced Materials}\ }\textbf {\bibinfo
  {volume} {21}},\ \bibinfo {pages} {4452} (\bibinfo {year}
  {2009})}\BibitemShut {NoStop}%
\bibitem [{\citenamefont {Yamauchi}\ \emph {et~al.}(2009)\citenamefont
  {Yamauchi}, \citenamefont {Fukushima},\ and\ \citenamefont
  {Picozzi}}]{Yamauchi2009}%
  \BibitemOpen
  \bibfield  {author} {\bibinfo {author} {\bibfnamefont {K.}~\bibnamefont
  {Yamauchi}}, \bibinfo {author} {\bibfnamefont {T.}~\bibnamefont
  {Fukushima}},\ and\ \bibinfo {author} {\bibfnamefont {S.}~\bibnamefont
  {Picozzi}},\ }\bibfield  {title} {\bibinfo {title} {{Ferroelectricity in
  multiferroic magnetite Fe3 O4 driven by noncentrosymmetric Fe2+ / Fe3+
  charge-ordering: First-principles study}},\ }\href
  {https://doi.org/10.1103/PhysRevB.79.212404} {\bibfield  {journal} {\bibinfo
  {journal} {Physical Review B - Condensed Matter and Materials Physics}\
  }\textbf {\bibinfo {volume} {79}},\ \bibinfo {pages} {212404} (\bibinfo
  {year} {2009})}\BibitemShut {NoStop}%
\bibitem [{\citenamefont {Senn}\ \emph {et~al.}(2012)\citenamefont {Senn},
  \citenamefont {Wright},\ and\ \citenamefont {Attfield}}]{Senn2012}%
  \BibitemOpen
  \bibfield  {author} {\bibinfo {author} {\bibfnamefont {M.~S.}\ \bibnamefont
  {Senn}}, \bibinfo {author} {\bibfnamefont {J.~P.}\ \bibnamefont {Wright}},\
  and\ \bibinfo {author} {\bibfnamefont {J.~P.}\ \bibnamefont {Attfield}},\
  }\bibfield  {title} {\bibinfo {title} {{Charge order and three-site
  distortions in the Verwey structure of magnetite}},\ }\href
  {https://doi.org/10.1038/nature10704} {\bibfield  {journal} {\bibinfo
  {journal} {Nature}\ }\textbf {\bibinfo {volume} {481}},\ \bibinfo {pages}
  {173} (\bibinfo {year} {2012})}\BibitemShut {NoStop}%
\bibitem [{\citenamefont {Radaelli}\ \emph {et~al.}(1997)\citenamefont
  {Radaelli}, \citenamefont {Cox},\ and\ \citenamefont
  {Marezio}}]{Radaelli1997}%
  \BibitemOpen
  \bibfield  {author} {\bibinfo {author} {\bibfnamefont {P.}~\bibnamefont
  {Radaelli}}, \bibinfo {author} {\bibfnamefont {D.}~\bibnamefont {Cox}},\ and\
  \bibinfo {author} {\bibfnamefont {M.}~\bibnamefont {Marezio}},\ }\bibfield
  {title} {\bibinfo {title} {{Charge, orbital, and magnetic ordering ins}},\
  }\href {https://doi.org/10.1103/PhysRevB.55.3015} {\bibfield  {journal}
  {\bibinfo  {journal} {Physical Review B - Condensed Matter and Materials
  Physics}\ }\textbf {\bibinfo {volume} {55}},\ \bibinfo {pages} {3015}
  (\bibinfo {year} {1997})}\BibitemShut {NoStop}%
\bibitem [{\citenamefont {Efremov}\ \emph {et~al.}(2004)\citenamefont
  {Efremov}, \citenamefont {{Van Den Brink}},\ and\ \citenamefont
  {Khomskii}}]{Efremov2004}%
  \BibitemOpen
  \bibfield  {author} {\bibinfo {author} {\bibfnamefont {D.~V.}\ \bibnamefont
  {Efremov}}, \bibinfo {author} {\bibfnamefont {J.}~\bibnamefont {{Van Den
  Brink}}},\ and\ \bibinfo {author} {\bibfnamefont {D.~I.}\ \bibnamefont
  {Khomskii}},\ }\bibfield  {title} {\bibinfo {title} {{Bond- Versus
  site-centred ordering and possible ferroelectricity in manganites}},\ }\href
  {https://doi.org/10.1038/nmat1236} {\bibfield  {journal} {\bibinfo  {journal}
  {Nature Materials}\ }\textbf {\bibinfo {volume} {3}},\ \bibinfo {pages} {853}
  (\bibinfo {year} {2004})}\BibitemShut {NoStop}%
\bibitem [{\citenamefont {Giovannetti}\ \emph
  {et~al.}(2009{\natexlab{b}})\citenamefont {Giovannetti}, \citenamefont
  {Kumar}, \citenamefont {{Van Den Brink}},\ and\ \citenamefont
  {Picozzi}}]{Giovannetti2009}%
  \BibitemOpen
  \bibfield  {author} {\bibinfo {author} {\bibfnamefont {G.}~\bibnamefont
  {Giovannetti}}, \bibinfo {author} {\bibfnamefont {S.}~\bibnamefont {Kumar}},
  \bibinfo {author} {\bibfnamefont {J.}~\bibnamefont {{Van Den Brink}}},\ and\
  \bibinfo {author} {\bibfnamefont {S.}~\bibnamefont {Picozzi}},\ }\bibfield
  {title} {\bibinfo {title} {{Magnetically induced electronic ferroelectricity
  in half-doped manganites}},\ }\href
  {https://doi.org/10.1103/PhysRevLett.103.037601} {\bibfield  {journal}
  {\bibinfo  {journal} {Physical Review Letters}\ }\textbf {\bibinfo {volume}
  {103}},\ \bibinfo {pages} {037601} (\bibinfo {year}
  {2009}{\natexlab{b}})}\BibitemShut {NoStop}%
\bibitem [{\citenamefont {Serrao}\ \emph {et~al.}(2007)\citenamefont {Serrao},
  \citenamefont {Sundaresan},\ and\ \citenamefont {Rao}}]{Serrao2007}%
  \BibitemOpen
  \bibfield  {author} {\bibinfo {author} {\bibfnamefont {C.~R.}\ \bibnamefont
  {Serrao}}, \bibinfo {author} {\bibfnamefont {A.}~\bibnamefont {Sundaresan}},\
  and\ \bibinfo {author} {\bibfnamefont {C.~N.}\ \bibnamefont {Rao}},\
  }\bibfield  {title} {\bibinfo {title} {{Multiferroic nature of charge-ordered
  rare earth manganites}},\ }\href
  {https://doi.org/10.1088/0953-8984/19/49/496217} {\bibfield  {journal}
  {\bibinfo  {journal} {Journal of Physics Condensed Matter}\ }\textbf
  {\bibinfo {volume} {19}},\ \bibinfo {pages} {496217} (\bibinfo {year}
  {2007})}\BibitemShut {NoStop}%
\bibitem [{\citenamefont {Yamauchi}\ and\ \citenamefont
  {Picozzi}(2013)}]{Yamauchi2013}%
  \BibitemOpen
  \bibfield  {author} {\bibinfo {author} {\bibfnamefont {K.}~\bibnamefont
  {Yamauchi}}\ and\ \bibinfo {author} {\bibfnamefont {S.}~\bibnamefont
  {Picozzi}},\ }\bibfield  {title} {\bibinfo {title} {{Mechanism of
  ferroelectricity in half-doped manganites with pseudocubic and bilayer
  structure}},\ }\href {https://doi.org/10.7566/JPSJ.82.113703} {\bibfield
  {journal} {\bibinfo  {journal} {Journal of the Physical Society of Japan}\
  }\textbf {\bibinfo {volume} {82}},\ \bibinfo {pages} {113703} (\bibinfo
  {year} {2013})}\BibitemShut {NoStop}%
\bibitem [{\citenamefont {Sagayama}\ \emph {et~al.}(2014)\citenamefont
  {Sagayama}, \citenamefont {Toyoda}, \citenamefont {Sugimoto}, \citenamefont
  {Maeda}, \citenamefont {Yamada},\ and\ \citenamefont {Arima}}]{Sagayama2014}%
  \BibitemOpen
  \bibfield  {author} {\bibinfo {author} {\bibfnamefont {H.}~\bibnamefont
  {Sagayama}}, \bibinfo {author} {\bibfnamefont {S.}~\bibnamefont {Toyoda}},
  \bibinfo {author} {\bibfnamefont {K.}~\bibnamefont {Sugimoto}}, \bibinfo
  {author} {\bibfnamefont {Y.}~\bibnamefont {Maeda}}, \bibinfo {author}
  {\bibfnamefont {S.}~\bibnamefont {Yamada}},\ and\ \bibinfo {author}
  {\bibfnamefont {T.}~\bibnamefont {Arima}},\ }\bibfield  {title} {\bibinfo
  {title} {{Ferroelectricity driven by charge ordering in the A -site ordered
  perovskite manganite SmBaMn2 O6}},\ }\href
  {https://doi.org/10.1103/PhysRevB.90.241113} {\bibfield  {journal} {\bibinfo
  {journal} {Physical Review B - Condensed Matter and Materials Physics}\
  }\textbf {\bibinfo {volume} {90}},\ \bibinfo {pages} {241113R} (\bibinfo
  {year} {2014})}\BibitemShut {NoStop}%
\bibitem [{\citenamefont {Morikawa}\ \emph {et~al.}(2012)\citenamefont
  {Morikawa}, \citenamefont {Tsuda}, \citenamefont {Maeda}, \citenamefont
  {Yamada},\ and\ \citenamefont {Arima}}]{Morikawa2012}%
  \BibitemOpen
  \bibfield  {author} {\bibinfo {author} {\bibfnamefont {D.}~\bibnamefont
  {Morikawa}}, \bibinfo {author} {\bibfnamefont {K.}~\bibnamefont {Tsuda}},
  \bibinfo {author} {\bibfnamefont {Y.}~\bibnamefont {Maeda}}, \bibinfo
  {author} {\bibfnamefont {S.}~\bibnamefont {Yamada}},\ and\ \bibinfo {author}
  {\bibfnamefont {T.~H.}\ \bibnamefont {Arima}},\ }\bibfield  {title} {\bibinfo
  {title} {{Charge and orbital order patterns in an a-site ordered
  perovskite-type manganite SmBaMn 2O 6 determined by convergent-beam electron
  diffraction}},\ }\href {https://doi.org/10.1143/JPSJ.81.093602} {\bibfield
  {journal} {\bibinfo  {journal} {Journal of the Physical Society of Japan}\
  }\textbf {\bibinfo {volume} {81}},\ \bibinfo {pages} {093602} (\bibinfo
  {year} {2012})}\BibitemShut {NoStop}%
\bibitem [{\citenamefont {Tang}\ \emph {et~al.}(2010)\citenamefont {Tang},
  \citenamefont {Hou}, \citenamefont {Zhang}, \citenamefont {Dong},\ and\
  \citenamefont {Shu}}]{Tang2010}%
  \BibitemOpen
  \bibfield  {author} {\bibinfo {author} {\bibfnamefont {M.~H.}\ \bibnamefont
  {Tang}}, \bibinfo {author} {\bibfnamefont {J.~W.}\ \bibnamefont {Hou}},
  \bibinfo {author} {\bibfnamefont {J.}~\bibnamefont {Zhang}}, \bibinfo
  {author} {\bibfnamefont {G.~J.}\ \bibnamefont {Dong}},\ and\ \bibinfo
  {author} {\bibfnamefont {W.}~\bibnamefont {Shu}},\ }\bibfield  {title}
  {\bibinfo {title} {{The giant dielectric tunability effect in bulk La2NiMnO 6
  around room temperature}},\ }\href
  {https://doi.org/10.1016/j.ssc.2010.05.029} {\bibfield  {journal} {\bibinfo
  {journal} {Solid State Communications}\ }\textbf {\bibinfo {volume} {150}},\
  \bibinfo {pages} {1453} (\bibinfo {year} {2010})}\BibitemShut {NoStop}%
\bibitem [{\citenamefont {Tang}\ \emph {et~al.}(2011)\citenamefont {Tang},
  \citenamefont {Xiao}, \citenamefont {Jiang}, \citenamefont {Hou},
  \citenamefont {Li},\ and\ \citenamefont {He}}]{Tang2011}%
  \BibitemOpen
  \bibfield  {author} {\bibinfo {author} {\bibfnamefont {M.~H.}\ \bibnamefont
  {Tang}}, \bibinfo {author} {\bibfnamefont {Y.~G.}\ \bibnamefont {Xiao}},
  \bibinfo {author} {\bibfnamefont {B.}~\bibnamefont {Jiang}}, \bibinfo
  {author} {\bibfnamefont {J.~W.}\ \bibnamefont {Hou}}, \bibinfo {author}
  {\bibfnamefont {J.~C.}\ \bibnamefont {Li}},\ and\ \bibinfo {author}
  {\bibfnamefont {J.}~\bibnamefont {He}},\ }\bibfield  {title} {\bibinfo
  {title} {{The giant dielectric tunability effect in bulk Y 2NiMnO 6 around
  room temperature}},\ }\href {https://doi.org/10.1007/s00339-011-6608-5}
  {\bibfield  {journal} {\bibinfo  {journal} {Applied Physics A: Materials
  Science and Processing}\ }\textbf {\bibinfo {volume} {105}},\ \bibinfo
  {pages} {679} (\bibinfo {year} {2011})}\BibitemShut {NoStop}%
\bibitem [{\citenamefont {Rout}\ and\ \citenamefont
  {Srinivasan}(2018)}]{Rout2018}%
  \BibitemOpen
  \bibfield  {author} {\bibinfo {author} {\bibfnamefont {P.~C.}\ \bibnamefont
  {Rout}}\ and\ \bibinfo {author} {\bibfnamefont {V.}~\bibnamefont
  {Srinivasan}},\ }\bibfield  {title} {\bibinfo {title} {{Epitaxial strain
  control of hole-doping induced phases in a multiferroic Mott insulator
  Bi2FeCrO6}},\ }\href {https://doi.org/10.1103/PhysRevLett.123.107201}
  {\bibfield  {journal} {\bibinfo  {journal} {Physical Review Letters}\
  }\textbf {\bibinfo {volume} {123}},\ \bibinfo {pages} {107201} (\bibinfo
  {year} {2018})}\BibitemShut {NoStop}%
\bibitem [{\citenamefont {He}\ and\ \citenamefont {Jin}(2016)}]{He2016a}%
  \BibitemOpen
  \bibfield  {author} {\bibinfo {author} {\bibfnamefont {X.}~\bibnamefont
  {He}}\ and\ \bibinfo {author} {\bibfnamefont {K.~J.}\ \bibnamefont {Jin}},\
  }\bibfield  {title} {\bibinfo {title} {{Engineering charge ordering into
  multiferroicity}},\ }\href@noop {} {\bibfield  {journal} {\bibinfo  {journal}
  {Physical Review B}\ }\textbf {\bibinfo {volume} {93}},\ \bibinfo {pages}
  {161108(R)} (\bibinfo {year} {2016})}\BibitemShut {NoStop}%
\bibitem [{\citenamefont {Park}\ \emph {et~al.}(2017)\citenamefont {Park},
  \citenamefont {Kumar},\ and\ \citenamefont {Rabe}}]{Park2017}%
  \BibitemOpen
  \bibfield  {author} {\bibinfo {author} {\bibfnamefont {S.~Y.}\ \bibnamefont
  {Park}}, \bibinfo {author} {\bibfnamefont {A.}~\bibnamefont {Kumar}},\ and\
  \bibinfo {author} {\bibfnamefont {K.~M.}\ \bibnamefont {Rabe}},\ }\bibfield
  {title} {\bibinfo {title} {{Charge-Order-Induced Ferroelectricity in
  LaVO3/SrVO3 Superlattices}},\ }\href@noop {} {\bibfield  {journal} {\bibinfo
  {journal} {Physical Review Letters}\ }\textbf {\bibinfo {volume} {118}},\
  \bibinfo {pages} {087602} (\bibinfo {year} {2017})}\BibitemShut {NoStop}%
\bibitem [{\citenamefont {Park}(2022)}]{Park2022}%
  \BibitemOpen
  \bibfield  {author} {\bibinfo {author} {\bibfnamefont {S.~Y.}\ \bibnamefont
  {Park}},\ }\bibfield  {title} {\bibinfo {title} {{First-principles study of
  charge-ordered insulating phases in Ruddlesden-Popper LaSr2V2O7}},\ }\href
  {https://doi.org/10.1016/j.cap.2022.09.009} {\bibfield  {journal} {\bibinfo
  {journal} {Current Applied Physics}\ }\textbf {\bibinfo {volume} {44}},\
  \bibinfo {pages} {110} (\bibinfo {year} {2022})}\BibitemShut {NoStop}%
\bibitem [{\citenamefont {Chang}\ \emph {et~al.}(2011)\citenamefont {Chang},
  \citenamefont {Jeng}, \citenamefont {Ren},\ and\ \citenamefont
  {Hsue}}]{Chang2011}%
  \BibitemOpen
  \bibfield  {author} {\bibinfo {author} {\bibfnamefont {T.~R.}\ \bibnamefont
  {Chang}}, \bibinfo {author} {\bibfnamefont {H.~T.}\ \bibnamefont {Jeng}},
  \bibinfo {author} {\bibfnamefont {C.~Y.}\ \bibnamefont {Ren}},\ and\ \bibinfo
  {author} {\bibfnamefont {C.~S.}\ \bibnamefont {Hsue}},\ }\bibfield  {title}
  {\bibinfo {title} {{Charge-orbital ordering and ferroelectric polarization in
  multiferroic TbMn2O5 from first principles}},\ }\href@noop {} {\bibfield
  {journal} {\bibinfo  {journal} {Physical Review B - Condensed Matter and
  Materials Physics}\ }\textbf {\bibinfo {volume} {84}},\ \bibinfo {pages}
  {024421} (\bibinfo {year} {2011})}\BibitemShut {NoStop}%
\bibitem [{\citenamefont {Wang}\ \emph {et~al.}(2007)\citenamefont {Wang},
  \citenamefont {Guo},\ and\ \citenamefont {He}}]{Wang2007}%
  \BibitemOpen
  \bibfield  {author} {\bibinfo {author} {\bibfnamefont {C.}~\bibnamefont
  {Wang}}, \bibinfo {author} {\bibfnamefont {G.~C.}\ \bibnamefont {Guo}},\ and\
  \bibinfo {author} {\bibfnamefont {L.}~\bibnamefont {He}},\ }\bibfield
  {title} {\bibinfo {title} {{Ferroelectricity driven by the noncentrosymmetric
  magnetic ordering in multiferroic TbMn2O5: A first-principles study}},\
  }\href@noop {} {\bibfield  {journal} {\bibinfo  {journal} {Physical Review
  Letters}\ }\textbf {\bibinfo {volume} {99}},\ \bibinfo {pages} {177202}
  (\bibinfo {year} {2007})}\BibitemShut {NoStop}%
\bibitem [{\citenamefont {Okamoto}\ \emph {et~al.}(2007)\citenamefont
  {Okamoto}, \citenamefont {Huang}, \citenamefont {Mou}, \citenamefont {Chao},
  \citenamefont {Lin}, \citenamefont {Park}, \citenamefont {Cheong},\ and\
  \citenamefont {Chen}}]{Okamoto2007}%
  \BibitemOpen
  \bibfield  {author} {\bibinfo {author} {\bibfnamefont {J.}~\bibnamefont
  {Okamoto}}, \bibinfo {author} {\bibfnamefont {D.~J.}\ \bibnamefont {Huang}},
  \bibinfo {author} {\bibfnamefont {C.~Y.}\ \bibnamefont {Mou}}, \bibinfo
  {author} {\bibfnamefont {K.~S.}\ \bibnamefont {Chao}}, \bibinfo {author}
  {\bibfnamefont {H.~J.}\ \bibnamefont {Lin}}, \bibinfo {author} {\bibfnamefont
  {S.}~\bibnamefont {Park}}, \bibinfo {author} {\bibfnamefont {S.~W.}\
  \bibnamefont {Cheong}},\ and\ \bibinfo {author} {\bibfnamefont {C.~T.}\
  \bibnamefont {Chen}},\ }\bibfield  {title} {\bibinfo {title} {{Symmetry of
  multiferroicity in a frustrated magnet TbMn2O5}},\ }\href
  {https://doi.org/10.1103/PhysRevLett.98.157202} {\bibfield  {journal}
  {\bibinfo  {journal} {Physical Review Letters}\ }\textbf {\bibinfo {volume}
  {98}},\ \bibinfo {pages} {1} (\bibinfo {year} {2007})}\BibitemShut {NoStop}%
\bibitem [{\citenamefont {{De Groot}}\ \emph {et~al.}(2012)\citenamefont {{De
  Groot}}, \citenamefont {Mueller}, \citenamefont {Rosenberg}, \citenamefont
  {Keavney}, \citenamefont {Islam}, \citenamefont {Kim},\ and\ \citenamefont
  {Angst}}]{deGroot2012}%
  \BibitemOpen
  \bibfield  {author} {\bibinfo {author} {\bibfnamefont {J.}~\bibnamefont {{De
  Groot}}}, \bibinfo {author} {\bibfnamefont {T.}~\bibnamefont {Mueller}},
  \bibinfo {author} {\bibfnamefont {R.~A.}\ \bibnamefont {Rosenberg}}, \bibinfo
  {author} {\bibfnamefont {D.~J.}\ \bibnamefont {Keavney}}, \bibinfo {author}
  {\bibfnamefont {Z.}~\bibnamefont {Islam}}, \bibinfo {author} {\bibfnamefont
  {J.~W.}\ \bibnamefont {Kim}},\ and\ \bibinfo {author} {\bibfnamefont
  {M.}~\bibnamefont {Angst}},\ }\bibfield  {title} {\bibinfo {title} {{Charge
  order in LuFe 2O 4: An unlikely route to ferroelectricity}},\ }\href
  {https://doi.org/10.1103/PhysRevLett.108.187601} {\bibfield  {journal}
  {\bibinfo  {journal} {Physical Review Letters}\ }\textbf {\bibinfo {volume}
  {108}},\ \bibinfo {pages} {187601} (\bibinfo {year} {2012})}\BibitemShut
  {NoStop}%
\bibitem [{\citenamefont {Ikeda}\ \emph {et~al.}(2005)\citenamefont {Ikeda},
  \citenamefont {Ohsumi}, \citenamefont {Ohwada}, \citenamefont {Ishii},
  \citenamefont {Inami}, \citenamefont {Kakurai}, \citenamefont {Murakami},
  \citenamefont {Yoshii}, \citenamefont {Mori}, \citenamefont {Horibe},\ and\
  \citenamefont {Kit{\^{o}}}}]{Ikeda2005}%
  \BibitemOpen
  \bibfield  {author} {\bibinfo {author} {\bibfnamefont {N.}~\bibnamefont
  {Ikeda}}, \bibinfo {author} {\bibfnamefont {H.}~\bibnamefont {Ohsumi}},
  \bibinfo {author} {\bibfnamefont {K.}~\bibnamefont {Ohwada}}, \bibinfo
  {author} {\bibfnamefont {K.}~\bibnamefont {Ishii}}, \bibinfo {author}
  {\bibfnamefont {T.}~\bibnamefont {Inami}}, \bibinfo {author} {\bibfnamefont
  {K.}~\bibnamefont {Kakurai}}, \bibinfo {author} {\bibfnamefont
  {Y.}~\bibnamefont {Murakami}}, \bibinfo {author} {\bibfnamefont
  {K.}~\bibnamefont {Yoshii}}, \bibinfo {author} {\bibfnamefont
  {S.}~\bibnamefont {Mori}}, \bibinfo {author} {\bibfnamefont {Y.}~\bibnamefont
  {Horibe}},\ and\ \bibinfo {author} {\bibfnamefont {H.}~\bibnamefont
  {Kit{\^{o}}}},\ }\bibfield  {title} {\bibinfo {title} {{Ferroelectricity from
  iron valence ordering in the charge-frustrated system LuFe2O4}},\ }\href
  {https://doi.org/10.1038/nature04039} {\bibfield  {journal} {\bibinfo
  {journal} {Nature}\ }\textbf {\bibinfo {volume} {436}},\ \bibinfo {pages}
  {1136} (\bibinfo {year} {2005})}\BibitemShut {NoStop}%
\bibitem [{\citenamefont {Angst}(2013)}]{Angst2013}%
  \BibitemOpen
  \bibfield  {author} {\bibinfo {author} {\bibfnamefont {M.}~\bibnamefont
  {Angst}},\ }\bibfield  {title} {\bibinfo {title} {{Ferroelectricity from iron
  valence ordering in rare earth ferrites?}},\ }\href
  {https://doi.org/10.1002/pssr.201307103} {\bibfield  {journal} {\bibinfo
  {journal} {Physica Status Solidi - Rapid Research Letters}\ }\textbf
  {\bibinfo {volume} {7}},\ \bibinfo {pages} {383} (\bibinfo {year}
  {2013})}\BibitemShut {NoStop}%
\bibitem [{\citenamefont {Balachandran}\ \emph {et~al.}(2018)\citenamefont
  {Balachandran}, \citenamefont {Emery}, \citenamefont {Gubernatis},
  \citenamefont {Lookman}, \citenamefont {Wolverton},\ and\ \citenamefont
  {Zunger}}]{Balachandran2018}%
  \BibitemOpen
  \bibfield  {author} {\bibinfo {author} {\bibfnamefont {P.~V.}\ \bibnamefont
  {Balachandran}}, \bibinfo {author} {\bibfnamefont {A.~A.}\ \bibnamefont
  {Emery}}, \bibinfo {author} {\bibfnamefont {J.~E.}\ \bibnamefont
  {Gubernatis}}, \bibinfo {author} {\bibfnamefont {T.}~\bibnamefont {Lookman}},
  \bibinfo {author} {\bibfnamefont {C.}~\bibnamefont {Wolverton}},\ and\
  \bibinfo {author} {\bibfnamefont {A.}~\bibnamefont {Zunger}},\ }\bibfield
  {title} {\bibinfo {title} {{Predictions of new AB O3 perovskite compounds by
  combining machine learning and density functional theory}},\ }\href@noop {}
  {\bibfield  {journal} {\bibinfo  {journal} {Physical Review Materials}\
  }\textbf {\bibinfo {volume} {2}},\ \bibinfo {pages} {043802} (\bibinfo {year}
  {2018})}\BibitemShut {NoStop}%
\bibitem [{\citenamefont {Lin}\ \emph {et~al.}(2009)\citenamefont {Lin},
  \citenamefont {Chen},\ and\ \citenamefont {Liu}}]{Lin2009}%
  \BibitemOpen
  \bibfield  {author} {\bibinfo {author} {\bibfnamefont {Y.~Q.}\ \bibnamefont
  {Lin}}, \bibinfo {author} {\bibfnamefont {X.~M.}\ \bibnamefont {Chen}},\ and\
  \bibinfo {author} {\bibfnamefont {X.~Q.}\ \bibnamefont {Liu}},\ }\bibfield
  {title} {\bibinfo {title} {{Relaxor-like dielectric behavior in La2NiMnO6
  double perovskite ceramics}},\ }\href
  {https://doi.org/10.1016/j.ssc.2009.02.028} {\bibfield  {journal} {\bibinfo
  {journal} {Solid State Communications}\ }\textbf {\bibinfo {volume} {149}},\
  \bibinfo {pages} {784} (\bibinfo {year} {2009})}\BibitemShut {NoStop}%
\bibitem [{\citenamefont {Qi}\ and\ \citenamefont {Rabe}(2022)}]{Qi2022b}%
  \BibitemOpen
  \bibfield  {author} {\bibinfo {author} {\bibfnamefont {Y.}~\bibnamefont
  {Qi}}\ and\ \bibinfo {author} {\bibfnamefont {K.~M.}\ \bibnamefont {Rabe}},\
  }\bibfield  {title} {\bibinfo {title} {{Electron-lattice coupling effects in
  nonadiabatic polarization switching of charge-order-induced
  ferroelectrics}},\ }\href {https://doi.org/10.1103/PhysRevB.106.125131}
  {\bibfield  {journal} {\bibinfo  {journal} {Physical Review B}\ }\textbf
  {\bibinfo {volume} {106}},\ \bibinfo {pages} {1} (\bibinfo {year}
  {2022})}\BibitemShut {NoStop}%
\bibitem [{\citenamefont {Rao}\ \emph {et~al.}(1998)\citenamefont {Rao},
  \citenamefont {Arulraj}, \citenamefont {Santosh},\ and\ \citenamefont
  {Cheetham}}]{Rao1998}%
  \BibitemOpen
  \bibfield  {author} {\bibinfo {author} {\bibfnamefont {C.~N.}\ \bibnamefont
  {Rao}}, \bibinfo {author} {\bibfnamefont {A.}~\bibnamefont {Arulraj}},
  \bibinfo {author} {\bibfnamefont {P.~N.}\ \bibnamefont {Santosh}},\ and\
  \bibinfo {author} {\bibfnamefont {A.~K.}\ \bibnamefont {Cheetham}},\
  }\bibfield  {title} {\bibinfo {title} {{Charge-Ordering in Manganates}},\
  }\href {https://doi.org/10.1021/cm980318e} {\bibfield  {journal} {\bibinfo
  {journal} {Chemistry of Materials}\ }\textbf {\bibinfo {volume} {10}},\
  \bibinfo {pages} {2714} (\bibinfo {year} {1998})}\BibitemShut {NoStop}%
\bibitem [{\citenamefont {Zhu}\ \emph {et~al.}(2022)\citenamefont {Zhu},
  \citenamefont {Kang},\ and\ \citenamefont {Parrish}}]{Zhu2022}%
  \BibitemOpen
  \bibfield  {author} {\bibinfo {author} {\bibfnamefont {Q.}~\bibnamefont
  {Zhu}}, \bibinfo {author} {\bibfnamefont {B.}~\bibnamefont {Kang}},\ and\
  \bibinfo {author} {\bibfnamefont {K.}~\bibnamefont {Parrish}},\ }\bibfield
  {title} {\bibinfo {title} {{Symmetry relation database and its application to
  ferroelectric materials discovery}},\ }\href
  {https://doi.org/10.1557/s43579-022-00268-4} {\bibfield  {journal} {\bibinfo
  {journal} {MRS Communications}\ }\textbf {\bibinfo {volume} {12}},\ \bibinfo
  {pages} {686} (\bibinfo {year} {2022})}\BibitemShut {NoStop}%
\bibitem [{\citenamefont {Abrahams}(1989)}]{Abrahams1989}%
  \BibitemOpen
  \bibfield  {author} {\bibinfo {author} {\bibfnamefont {S.~C.}\ \bibnamefont
  {Abrahams}},\ }\bibfield  {title} {\bibinfo {title} {{Structurally based
  predictions of ferroelectricity in seven inorganic materials with space group
  Pba2 and two experimental confirmations}},\ }\href
  {https://doi.org/https://doi.org/10.1107/S0108768189001072} {\bibfield
  {journal} {\bibinfo  {journal} {Acta Crystallographica Section B}\ }\textbf
  {\bibinfo {volume} {45}},\ \bibinfo {pages} {228} (\bibinfo {year}
  {1989})}\BibitemShut {NoStop}%
\bibitem [{\citenamefont {Smidt}\ \emph {et~al.}(2020)\citenamefont {Smidt},
  \citenamefont {Mack}, \citenamefont {Reyes-Lillo}, \citenamefont {Jain},\
  and\ \citenamefont {Neaton}}]{Smidt2020}%
  \BibitemOpen
  \bibfield  {author} {\bibinfo {author} {\bibfnamefont {T.~E.}\ \bibnamefont
  {Smidt}}, \bibinfo {author} {\bibfnamefont {S.~A.}\ \bibnamefont {Mack}},
  \bibinfo {author} {\bibfnamefont {S.~E.}\ \bibnamefont {Reyes-Lillo}},
  \bibinfo {author} {\bibfnamefont {A.}~\bibnamefont {Jain}},\ and\ \bibinfo
  {author} {\bibfnamefont {J.~B.}\ \bibnamefont {Neaton}},\ }\bibfield  {title}
  {\bibinfo {title} {{An automatically curated first- principles database of
  ferroelectrics}},\ }\href@noop {} {\bibfield  {journal} {\bibinfo  {journal}
  {Scientific Data}\ }\textbf {\bibinfo {volume} {7}},\ \bibinfo {pages} {72}
  (\bibinfo {year} {2020})}\BibitemShut {NoStop}%
\bibitem [{\citenamefont {Ricci}\ \emph {et~al.}(2024)\citenamefont {Ricci},
  \citenamefont {Reyes-Lillo}, \citenamefont {Mack},\ and\ \citenamefont
  {Neaton}}]{Ricci2024}%
  \BibitemOpen
  \bibfield  {author} {\bibinfo {author} {\bibfnamefont {F.}~\bibnamefont
  {Ricci}}, \bibinfo {author} {\bibfnamefont {S.~E.}\ \bibnamefont
  {Reyes-Lillo}}, \bibinfo {author} {\bibfnamefont {S.~A.}\ \bibnamefont
  {Mack}},\ and\ \bibinfo {author} {\bibfnamefont {J.~B.}\ \bibnamefont
  {Neaton}},\ }\bibfield  {title} {\bibinfo {title} {{Candidate ferroelectrics
  via ab initio high-throughput screening of polar materials}},\ }\href
  {https://doi.org/10.1038/s41524-023-01193-3} {\bibfield  {journal} {\bibinfo
  {journal} {npj Computational Materials}\ }\textbf {\bibinfo {volume} {10}},\
  \bibinfo {pages} {1} (\bibinfo {year} {2024})}\BibitemShut {NoStop}%
\bibitem [{\citenamefont {Forero-Correa}\ \emph {et~al.}(2025)\citenamefont
  {Forero-Correa}, \citenamefont {Varas-Salinas}, \citenamefont {Castillo},\
  and\ \citenamefont {Reyes-Lillo}}]{Forero-Correa2025}%
  \BibitemOpen
  \bibfield  {author} {\bibinfo {author} {\bibfnamefont {N.}~\bibnamefont
  {Forero-Correa}}, \bibinfo {author} {\bibfnamefont {N.}~\bibnamefont
  {Varas-Salinas}}, \bibinfo {author} {\bibfnamefont {T.~M.}\ \bibnamefont
  {Castillo}},\ and\ \bibinfo {author} {\bibfnamefont {S.~E.}\ \bibnamefont
  {Reyes-Lillo}},\ }\bibfield  {title} {\bibinfo {title} {{Intrinsic inverse
  band gap versus polarization relation in ferroelectric materials}},\ }\href
  {https://doi.org/10.1103/PhysRevMaterials.9.034403} {\bibfield  {journal}
  {\bibinfo  {journal} {Physical Review Materials}\ }\textbf {\bibinfo {volume}
  {9}},\ \bibinfo {pages} {1} (\bibinfo {year} {2025})}\BibitemShut {NoStop}%
\bibitem [{sup()}]{supp}%
  \BibitemOpen
  \href@noop {} {}\bibinfo {note} {See Supplemental Material at
  URL-will-be-inserted-by-publisher.}\BibitemShut {Stop}%
\bibitem [{\citenamefont {Reyes-Lillo}\ \emph {et~al.}(2014)\citenamefont
  {Reyes-Lillo}, \citenamefont {Garrity},\ and\ \citenamefont
  {Rabe}}]{Reyes-Lillo2014}%
  \BibitemOpen
  \bibfield  {author} {\bibinfo {author} {\bibfnamefont {S.~E.}\ \bibnamefont
  {Reyes-Lillo}}, \bibinfo {author} {\bibfnamefont {K.~F.}\ \bibnamefont
  {Garrity}},\ and\ \bibinfo {author} {\bibfnamefont {K.~M.}\ \bibnamefont
  {Rabe}},\ }\bibfield  {title} {\bibinfo {title} {{Antiferroelectricity in
  thin-film ZrO2 from first principles}},\ }\href
  {https://doi.org/10.1103/PhysRevB.90.140103} {\bibfield  {journal} {\bibinfo
  {journal} {Physical Review B - Condensed Matter and Materials Physics}\
  }\textbf {\bibinfo {volume} {90}},\ \bibinfo {pages} {1} (\bibinfo {year}
  {2014})}\BibitemShut {NoStop}%
\bibitem [{\citenamefont {Horton}\ \emph {et~al.}(2019)\citenamefont {Horton},
  \citenamefont {Montoya}, \citenamefont {Liu},\ and\ \citenamefont
  {Persson}}]{Horton2019}%
  \BibitemOpen
  \bibfield  {author} {\bibinfo {author} {\bibfnamefont {M.~K.}\ \bibnamefont
  {Horton}}, \bibinfo {author} {\bibfnamefont {J.~H.}\ \bibnamefont {Montoya}},
  \bibinfo {author} {\bibfnamefont {M.}~\bibnamefont {Liu}},\ and\ \bibinfo
  {author} {\bibfnamefont {K.~A.}\ \bibnamefont {Persson}},\ }\bibfield
  {title} {\bibinfo {title} {{High-throughput prediction of the ground-state
  collinear magnetic order of inorganic materials using Density Functional
  Theory}},\ }\href@noop {} {\bibfield  {journal} {\bibinfo  {journal} {npj
  Computational Materials}\ }\textbf {\bibinfo {volume} {5}},\ \bibinfo {pages}
  {64} (\bibinfo {year} {2019})}\BibitemShut {NoStop}%
\bibitem [{\citenamefont {Wang}\ \emph {et~al.}(2014)\citenamefont {Wang},
  \citenamefont {Lee}, \citenamefont {Zhang}, \citenamefont {Shang},
  \citenamefont {Chen}, \citenamefont {Derecskei-Kovacs},\ and\ \citenamefont
  {Liu}}]{Wang2014}%
  \BibitemOpen
  \bibfield  {author} {\bibinfo {author} {\bibfnamefont {Y.}~\bibnamefont
  {Wang}}, \bibinfo {author} {\bibfnamefont {S.~H.}\ \bibnamefont {Lee}},
  \bibinfo {author} {\bibfnamefont {L.~A.}\ \bibnamefont {Zhang}}, \bibinfo
  {author} {\bibfnamefont {S.~L.}\ \bibnamefont {Shang}}, \bibinfo {author}
  {\bibfnamefont {L.~Q.}\ \bibnamefont {Chen}}, \bibinfo {author}
  {\bibfnamefont {A.}~\bibnamefont {Derecskei-Kovacs}},\ and\ \bibinfo {author}
  {\bibfnamefont {Z.~K.}\ \bibnamefont {Liu}},\ }\bibfield  {title} {\bibinfo
  {title} {{Quantifying charge ordering by density functional theory: Fe 3O4
  and CaFeO3}},\ }\href {https://doi.org/10.1016/j.cplett.2014.05.044}
  {\bibfield  {journal} {\bibinfo  {journal} {Chemical Physics Letters}\
  }\textbf {\bibinfo {volume} {607}},\ \bibinfo {pages} {81} (\bibinfo {year}
  {2014})}\BibitemShut {NoStop}%
\bibitem [{\citenamefont {Verwey}(1939)}]{Verwey1939}%
  \BibitemOpen
  \bibfield  {author} {\bibinfo {author} {\bibfnamefont {E.~J.~W.}\
  \bibnamefont {Verwey}},\ }\bibfield  {title} {\bibinfo {title} {{Electronic
  Conduction of Magnetite (Fe3O4) and its Transition Point at Low
  Temperatures}},\ }\href {https://doi.org/10.1351/goldbook.v06612} {\bibfield
  {journal} {\bibinfo  {journal} {Nature}\ }\textbf {\bibinfo {volume} {144}},\
  \bibinfo {pages} {327} (\bibinfo {year} {1939})}\BibitemShut {NoStop}%
\bibitem [{\citenamefont {Verdugo-Ihl}\ \emph {et~al.}(2020)\citenamefont
  {Verdugo-Ihl}, \citenamefont {Ciobanu}, \citenamefont {Cook}, \citenamefont
  {Ehrig}, \citenamefont {Slattery},\ and\ \citenamefont
  {Courtney-Davies}}]{Verdugo-Ihl2020}%
  \BibitemOpen
  \bibfield  {author} {\bibinfo {author} {\bibfnamefont {M.~R.}\ \bibnamefont
  {Verdugo-Ihl}}, \bibinfo {author} {\bibfnamefont {C.~L.}\ \bibnamefont
  {Ciobanu}}, \bibinfo {author} {\bibfnamefont {N.~J.}\ \bibnamefont {Cook}},
  \bibinfo {author} {\bibfnamefont {K.}~\bibnamefont {Ehrig}}, \bibinfo
  {author} {\bibfnamefont {A.}~\bibnamefont {Slattery}},\ and\ \bibinfo
  {author} {\bibfnamefont {L.}~\bibnamefont {Courtney-Davies}},\ }\bibfield
  {title} {\bibinfo {title} {{Trace-element remobilisation from W–Sn–U–Pb
  zoned hematite: Nanoscale insights into a mineral geochronometer behaviour
  during interaction with fluids}},\ }\href
  {https://doi.org/10.1180/mgm.2020.49} {\bibfield  {journal} {\bibinfo
  {journal} {Mineralogical Magazine}\ }\textbf {\bibinfo {volume} {84}},\
  \bibinfo {pages} {502} (\bibinfo {year} {2020})}\BibitemShut {NoStop}%
\bibitem [{\citenamefont {Iizumi}\ \emph {et~al.}(1982)\citenamefont {Iizumi},
  \citenamefont {Koetzle}, \citenamefont {Shirane}, \citenamefont {Chikazumi},
  \citenamefont {Matsui},\ and\ \citenamefont {Todo}}]{Iizumi1982}%
  \BibitemOpen
  \bibfield  {author} {\bibinfo {author} {\bibfnamefont {M.}~\bibnamefont
  {Iizumi}}, \bibinfo {author} {\bibfnamefont {T.~F.}\ \bibnamefont {Koetzle}},
  \bibinfo {author} {\bibfnamefont {G.}~\bibnamefont {Shirane}}, \bibinfo
  {author} {\bibfnamefont {S.}~\bibnamefont {Chikazumi}}, \bibinfo {author}
  {\bibfnamefont {M.}~\bibnamefont {Matsui}},\ and\ \bibinfo {author}
  {\bibfnamefont {S.}~\bibnamefont {Todo}},\ }\bibfield  {title} {\bibinfo
  {title} {{Structure of magnetite (Fe3O4) below the Verwey transition
  temperature.}},\ }\href {https://doi.org/10.1107/s0567740882008176}
  {\bibfield  {journal} {\bibinfo  {journal} {Acta Crystallographica}\ }\textbf
  {\bibinfo {volume} {B38}},\ \bibinfo {pages} {2121} (\bibinfo {year}
  {1982})}\BibitemShut {NoStop}%
\bibitem [{\citenamefont {Lauer}\ \emph {et~al.}(2004)\citenamefont {Lauer},
  \citenamefont {Valenti}, \citenamefont {Kandpal},\ and\ \citenamefont
  {Seshadri}}]{Lauer2004}%
  \BibitemOpen
  \bibfield  {author} {\bibinfo {author} {\bibfnamefont {M.}~\bibnamefont
  {Lauer}}, \bibinfo {author} {\bibfnamefont {R.}~\bibnamefont {Valenti}},
  \bibinfo {author} {\bibfnamefont {H.~C.}\ \bibnamefont {Kandpal}},\ and\
  \bibinfo {author} {\bibfnamefont {R.}~\bibnamefont {Seshadri}},\ }\bibfield
  {title} {\bibinfo {title} {{First-principles electronic structure of spinel
  LiCr2O4: A A possible half-metal}},\ }\href
  {https://doi.org/10.1103/PhysRevB.69.075117} {\bibfield  {journal} {\bibinfo
  {journal} {Physical Review B - Condensed Matter and Materials Physics}\
  }\textbf {\bibinfo {volume} {69}},\ \bibinfo {pages} {1} (\bibinfo {year}
  {2004})}\BibitemShut {NoStop}%
\bibitem [{\citenamefont {Rodr{\'{i}}guez-Carvajal}\ \emph
  {et~al.}(1998)\citenamefont {Rodr{\'{i}}guez-Carvajal}, \citenamefont
  {Rousse}, \citenamefont {Masquelier},\ and\ \citenamefont
  {Hervieu}}]{Rodriguez-Carvajal1998}%
  \BibitemOpen
  \bibfield  {author} {\bibinfo {author} {\bibfnamefont {J.}~\bibnamefont
  {Rodr{\'{i}}guez-Carvajal}}, \bibinfo {author} {\bibfnamefont
  {G.}~\bibnamefont {Rousse}}, \bibinfo {author} {\bibfnamefont
  {C.}~\bibnamefont {Masquelier}},\ and\ \bibinfo {author} {\bibfnamefont
  {M.}~\bibnamefont {Hervieu}},\ }\bibfield  {title} {\bibinfo {title}
  {{Electronic crystallization in a lithium battery material: Columnar ordering
  of electrons and holes in the spinel LiMn2O4}},\ }\href
  {https://doi.org/10.1103/PhysRevLett.81.4660} {\bibfield  {journal} {\bibinfo
   {journal} {Physical Review Letters}\ }\textbf {\bibinfo {volume} {81}},\
  \bibinfo {pages} {4660} (\bibinfo {year} {1998})}\BibitemShut {NoStop}%
\end{thebibliography}

%

%

\end{document}